\documentclass[floats,floatfix,amssymb,prl,twocolumn,superscriptaddress,nofootinbib]{revtex4-1}

\makeatletter
\def\l@subsubsection#1#2{}
\def\l@subsubsubsection#1#2{}
\makeatother

\usepackage{graphicx,amssymb,amsmath,amsthm,amsfonts,epsfig,epsf,fixmath}
\usepackage[usenames]{color}
\usepackage{epstopdf}

\usepackage{aas_macros}
\usepackage{bm}
\usepackage{dcolumn}
\usepackage{latexsym}
\usepackage{rotating}
\usepackage{longtable}

\usepackage{enumerate}
\usepackage{tensor,multirow}
\usepackage{url}
\usepackage[dvipsnames]{xcolor}
\usepackage[unicode]{hyperref}
\hypersetup{colorlinks=true, citecolor=MidnightBlue,
            linkcolor=MidnightBlue, urlcolor=MidnightBlue, linktocpage=true}
\usepackage{comment}

\newcommand{\sus}[1]{{\textcolor{red}{\sf{[SB: #1]}} }}

\newcommand{\tn}{\textnormal}

\newcommand{\dd}{\mathrm{d}}

\newcommand{\GSSI}{Gran Sasso Science Institute (GSSI), I-67100 L’Aquila, Italy}
\newcommand{\GranSasso}{INFN, Laboratori Nazionali del Gran Sasso, I-67100 Assergi, Italy}

\begin{document}

\title{Detecting massive scalar fields with Extreme Mass-Ratio Inspirals\\ }

\author{Susanna Barsanti}
\affiliation{Dipartimento di Fisica, ``Sapienza'' Universit\`a di Roma, Piazzale 
Aldo Moro 5, 00185, Roma, Italy}
\affiliation{Sezione INFN Roma1, Roma 00185, Italy}
\author{Andrea Maselli}
\affiliation{\GSSI}
\affiliation{\GranSasso}
\author{Thomas P. Sotiriou}
\affiliation{School of Mathematical Sciences \& School of Physics and Astronomy, University of Nottingham, University Park, Nottingham, NG7 2RD, UK}
\affiliation{Nottingham Centre of Gravity, University of Nottingham, University Park, Nottingham, NG7 2RD, UK}
\author{Leonardo Gualtieri}
\affiliation{Dipartimento di Fisica,  Universit\`a di Pisa  \& Sezione INFN Pisa, L. Bruno Pontecorvo
3, 56127 Pisa, Italy}

\begin{abstract} 
We study the imprint of light scalar fields on gravitational waves from extreme mass ratio inspirals --- binary systems with a very large mass asymmetry.  We first show that, to leading order in the mass ratio, any effects of the scalar on the waveform are captured fully by two parameters: the mass of the scalar and the scalar charge of the secondary compact object. We then use this theory-agnostic framework to show that the future observations by LISA will be able to {\em simultaneously} measure both of these parameters with enough accuracy to detect ultra-light scalars. 
\end{abstract}

\maketitle

%-----------------------------------------------------
\noindent{{\bf{\em Introduction.}}} 
Asymmetric binaries represent a new family of compact sources of gravitational waves (GWs) with an exceptional discovery potential. Mildly asymmetric binaries have already been observed by LVK \cite{LIGOScientific:2021djp}. LISA \cite{Audley:2017drz} 
is expected to observe compact binaries with much lower mass ratios, up to a factor of $10^5$.
These sources can lie in the detection band 
for years, rather than minutes, because in the final stages of the inspiral the evolution timescale of a highly asymmetric binary is proportional to the mass ratio. The large number of gravitational wave cycles produced while the smaller (secondary) object is performing relativistic orbits around the larger (primary) object are expected to offer unprecedented precision in parameter estimation for astrophysics \cite{Seoane:2021kkk,Laghi:2021pqk,Berry:2019wgg,McGee:2018qwb,Amaro-Seoane:2007osp,Cardoso:2019rou,Barausse:2014tra,Yunes:2011ws,Kocsis:2011dr,Destounis:2021mqv,Cardoso:2022whc,Cole:2022fir,Bamber:2022pbs} and fundamental physics\,\cite{Barack:2018yly,Barausse:2020rsu,Barausse:2016eii,Blazquez-Salcedo:2016enn,Glampedakis:2005cf,Barack:2006pq,Cardoso:2018zhm,Datta:2019epe,Pani:2019cyc,Maggio:2021uge,Destounis:2020kss,Piovano:2020ooe,Annulli:2020ilw,Sago:2021iku,Piovano:2022ojl} alike.

Extreme mass-ratio inspirals (EMRIs), in which a stellar-mass secondary of mass $m_p$ evolves around a supermassive black hole (BH)  of mass $M$ with mass ratios of $q=m_p/M\sim10^{-3}\div10^{-6}$, are perhaps the most promising sources in this respect. 
In particular, they can be a very sensitive probe of new fundamental scalar fields\,\cite{Pani:2011xj,Yunes:2011aa,Hannuksela:2018izj,Maselli:2020zgv,Barausse:2020rsu,Maselli:2021men,Collodel:2021jwi,Barsanti:2022ana}. 
Scalars are ubiquitous in cosmological models of dark energy and/or dark matter and in extension of the Standard Model or General Relativity \cite{Berti:2015itd,Barack:2018yly}.

Harnessing the potential of asymmetric binaries, and EMRIs in particular, for detecting or constraining new fundamental fields requires developing accurate waveforms in scenarios that include such fields. Remarkably, for massless scalars, it was shown in \cite{Maselli:2020zgv} that this can be done in a theory-agnostic way to leading order in the mass ratio. Any (self)interaction of the scalar that respects shift-symmetry --- the symmetry that protects the scalar from acquiring a mass --- affects the waveform only through a single parameter: the scalar charge per unit mass of the secondary, $d$. This framework was used in \cite{Maselli:2021men} to produce the first forecasts for LISA's ability to detect scalar charge.

It is worth stressing that massive scalars are expected to leave an observable imprint on compact objects only if their Compton wavelength, the inverse of their mass, is comparable to the size of the objects or the length-scales of their system. 
In geometrical $(G=c=1)$ units, if $M$ is the length-scale of the source (e.g., for a BH system, the BH mass) and $\mu_s\hbar$ is the scalar field mass, the condition is roughly $\mu_sM\lesssim1$.  
We note that\,\cite{Brito:2015oca}
\begin{equation}
\mu_s\,[{\rm eV}]\simeq\left(\frac{\mu_s M}{0.75}\right)\cdot \left(\frac{10^6 M_{\odot}}{M}\right)10^{-16}\,{\rm eV}\,. 
\end{equation}
Hence, the scalars that GW observations can currently probe would have 
masses smaller than $\sim10^{-16}$ eV ({\it ultra-light scalar fields}, see e.g.\,\cite{Brito:2017zvb} and references therein).

Nonetheless, the assumption of a strictly vanishing mass and of shift symmetry, as in \cite{Maselli:2020zgv}, can be too restrictive. Certain scenarios, such a superradiance-induced clouds \cite{Brito:2015oca} or scalarization \cite{Damour:1993hw,Silva:2017uqg,Doneva:2017bvd,Dima:2020yac,Doneva:2022ewd} rely of the presence of a mass or of interactions that violate shift symmetry to generate scalar charge. Moreover, measuring the mass of an ultra-light scalar is in itself an exciting prospect. Indeed, significant effort has already been put in constraining the mass of scalar fields using pulsar or LVK observations, {\em e.g.} \cite{Antoniadis:2013pzd,Yamada:2019zrb,Ramazanoglu:2016kul,Brito:2017zvb}. 
The main goal of this paper is to demonstrate that EMRIs are sensitive probes of ultra-light scalar fields, which can allow us to measure the scalar charge per unit mass of the secondary and the mass of the scalar field {\em simultaneously}, and with impressive precision.

\noindent{{\bf{\em Setup.}}} We consider the general action 
\begin{equation}
 S\left[\textbf{g}, \varphi, \Psi \right] = S_0\left[\textbf{g}, \varphi\right] + \alpha S_{\rm c} \left[\textbf{g}, \varphi\right] + S_{\rm m}\left[\textbf{g}, \varphi, \Psi\right]\ ,\label{eq:action}
 \end{equation}
where   
\begin{equation}
    S_0 = \int d^4 x \frac{\sqrt{-g}}{16 \pi} \left(R - \frac{1}{2} \partial_\mu \varphi \partial^{\mu} \varphi - \frac{1}{2} \mu_s^2 \varphi^2  \right)\,,
 \end{equation}
$R$ is the Ricci scalar, $\mu_s$ is the mass of the scalar field.  
$S_c$ encodes all additional interaction of the scalar field, including nonminimal couplings to gravity, and is assumed to be analytic in $\varphi$. $S_{\rm m}$ describes matter fields. In an EMRI the secondary object can be treated as a point particle, by replacing $S_{\rm m}$ with the ``skeletonized action''\,\cite{1975ApJ...196L..59E} 
\begin{equation}
S_p=-\int m(\varphi) \sqrt{g_{\alpha\beta}\frac{dy^{\alpha}_p}{d\lambda} \frac{dy^{\beta}_p}{d\lambda}}d\lambda\,,
\end{equation}
where $m(\varphi)$ is a scalar function.
By varying Eq.~\eqref{eq:action} with respect to $\textbf{g}$ and $\varphi$  we obtain the field equations 
\begin{align}
G_{\mu\nu} = &-\frac{16\pi\alpha}{\sqrt{-g}}\frac{\delta S_c}{\delta g^{\mu\nu}}+8 \pi T^{\rm scal}_{\mu \nu}+8 \pi T^{ p}_{\mu \nu} , \label{eq:fieldstens}\\
%&+8 \pi  \int m(\varphi) \frac{\delta^{(4)}\left(x-y_p(\lambda)\right)}{\sqrt{-g}}\frac{dy^{\alpha}_p}{d\lambda} \frac{dy^{\beta}_p}{d\lambda}d\lambda\ ,\label{eq:fieldstens}\\ 
( \square - \mu_s^2 ) \varphi =& -\frac{8\pi\alpha}{\sqrt{-g}}\frac{\delta S_c}{\delta \varphi}+ 16 \pi  \frac{\delta S_p}{\delta \varphi},\label{eq:fieldscal}
%&+ 16 \pi \int m'\left(\varphi\right) \frac{\delta ^{(4)}\left(x - y_p (\lambda)\right)}{\sqrt{-g}}d\lambda\ ,\label{eq:fieldscal}
\end{align}
where $T^{\rm scal}_{\mu\nu}$ is the standard scalar-field stress-energy tensor and $T^{p}_{\mu\nu}$ is the stress-energy tensor for $S_p$. 
Eqs.\,\eqref{eq:fieldstens},\,\eqref{eq:fieldscal} can be solved perturbatively in $q=m_p/M\ll1$, with the secondary acting as a perturbation of the massive BH background. 

We assume that $\alpha$ has negative mass dimensions in units where $c=\hbar=1$ ({\em i.e.} it suppresses irrelevant operators), or positive length dimensions in the $G=c=\hbar=1$ geometric units that we use here. Then following \cite{Maselli:2020zgv} one can relate $\alpha$ to $q$ as follows, $\alpha/M^n=(\alpha/m_p^n) q^n$, where $n$ is a positive integer. We further assume that $\alpha/m_p^n$ is not much larger than 1. 
%Provided that $\mu_s m_p < 1$ this 
This
assumption is justified by the fact that we have not already detected deviations from Kerr for black holes of a few solar masses or in weak field \cite{Nair:2019iur} (note that the statement above is correct as long as $\mu_s m_p < 1$, which is always the case for the light scalar fields considered in this paper).  
Finally, for $\alpha=0$, the theory we are considering is covered by no-hair theorems \cite{Bekenstein:1995un,Sotiriou:2011dz} and hence the primary would be a Kerr black hole with $\varphi=0$. Combining all of the above, one can treat the deviations from the Kerr metric and the EMRI dynamics perturbatively, with $q$ as a single book-keeping parameter. When $\mu_s M\ll1$, the mass of the scalar can be neglected and one recovers the results of \cite{Maselli:2020zgv}, while when $\mu_sM$ becomes $O(1)$ it is essential to include its contribution, as we do below. 

We will only consider quantities to leading order in the mass ratio. Hence 
$T^{\rm scal}_{\mu\nu}$ and $\delta S_c/\delta \varphi$, which are  quadratic in $q$, 
can be neglected. The scalar perturbation $\varphi_1$ is then fully determined by 
the secondary. 
In a buffer region close to the secondary, small enough to be inside its world-tube, 
but far away such that the metric can be considered as a perturbation of flat spacetime, 
Eq. \eqref{eq:fieldscal} reduces to
\begin{equation}
  ( \square - \mu_s^2 ) \varphi_1 = 0 \ , 
\end{equation}
whose solution,  in a reference frame $\{\tilde{x}_\mu\}$ centered on the particle, has  the form
\begin{equation}
   \varphi_1 \simeq \frac{m_{\rm p} d}{\tilde{r}} e^{-\mu_s\tilde{r}}+O\left(\frac{m_{\rm p}^2}{{\tilde r}^2}   e^{-\mu_s \tilde{r}} \right)   \, ,\label{eq:sollocal}
\end{equation}
where $d$ is the scalar charge of the secondary. By matching Eq.~\eqref{eq:sollocal} with the solution of Eq.~\eqref{eq:fieldscal} in the buffer region, we find that $m(0)=m_{\rm p}$ and $m'(0)/m(0)=-d/4$.

Eqs.\,\eqref{eq:fieldstens}, \eqref{eq:fieldscal} can then be written as:
\begin{align}
    G^{\alpha \beta} =&~ 8 \pi m_{\rm p}  \int \frac{\delta^{(4)}(x-y_p(\lambda))}{\sqrt{-g}}\frac{dy^{\alpha}_p}{d\lambda} \frac{dy^{\beta}_p}{d\lambda} d\lambda \ ,\label{eq:gr}\\
    \left(\square - \mu_s^2\right)\varphi =& - 4 \pi d m_{\rm p} \int \frac{\delta^{(4)} (x - y_p(\lambda)) }{\sqrt{-g}}d\lambda\ ,     \label{eq:scal}
\end{align}
where we have replaced explicit expressions for $T^{ p}_{\mu \nu}$ and $\delta S_p/\delta \varphi$,  and $y_p^\mu$ identifies the worldline followed by the secondary.
Eqs.\,\eqref{eq:gr},\,\eqref{eq:scal} are solved perturbatively following the Teukolsky approach \cite{Teukolsky:1973ha}. The scalar field is decomposed in spheroidal harmonics as a sum over multipoles $(\ell, m)$. Details on the scalar perturbations are given in Appendix~\ref{appendix:scalar perturbations}.

The total energy loss emitted by both the scalar and the gravitational sector is the sum of the contributions  at the horizon and at infinity:
\begin{equation}
    \dot{E}_\tn{GW} =  \sum_{i=+,-}[\dot{E}^{i}_{\rm grav}+\dot{E}^{i}_{\rm scal}]\ =  \dot{E}_\tn{grav} +  \dot{E}_\tn{scal}\ , \label{totflux}
\end{equation}
where the dot indicates the time derivative.
Since the source term of the scalar field equation depends linearly on 
the charge, the scalar energy flux can be written as $\dot{E}_\tn{scal} = d^2 \dot{\bar{E}}_\tn{scal}$, such that $q^{-2}\dot{\bar{E}}_\tn{scal}$ only depends 
on ($r/M$, $a/M$, $\mu_s M$).

The flux at infinity identically vanishes for frequencies smaller than the scalar field mass, $\omega < \mu_s$. This is a typical behaviour for massive scalar fields (see e.g.\,\cite{Alsing:2011er,Berti:2012bp,Ramazanoglu:2016kul}.
Therefore, for every combination of $(\ell, m)$ a specific radius $r_\textnormal{s}$ exists such that for $r > r_\textnormal{s}$ the energy flux at infinity vanishes. The general behaviour of the scalar energy flux as a function of the orbital radius and of $\mu_s$ is discussed in Appendix \ref{app:fluxes}. 

Unlike the emission at infinity, the flux at horizon is present for each value of the orbital frequency, and contributes to the binary's orbital evolution throughout the entire inspiral. Moreover it shows a new important feature, the appearence of resonances, which are not present if the scalar field is massless. 
Resonances occur when the binary orbital frequencies are comparable with those of the scalar quasi-normal modes of the BH background spacetime. 
In this case the energy emission grows towards a peak which can be either positive or negative depending on the BH spin and on the superradiance condition $\omega<m\Omega_h$, where $\Omega_h = a/(2Mr_h)$ and $r_h = M + \sqrt{M^2 - a^2}$. If the peak is negative the scalar radiation can be strong enough to counterbalance the gravitational emission, giving rise to floating orbits \cite{Yunes:2011aa, Cardoso:2011xi}. Determining whether floating orbits persist at post-adiabatic level or how quickly the secondary moves through a resonance requires self-force calculations \cite{Yunes:2011aa}, which are beyond the scope of this paper. Hereafter, we neglect resonances, which is a rather conservative approach. Taking them into account is expected to make the waveform more distinguishable from a EMRI waveform in GR and hence improve parameter estimation and our ability to detect a new scalar.

The gravitational and the massless scalar fluxes have been computed by making use of the Black Hole Perturbation Toolkit \cite{BHPToolkit}  while for the massive scalar fluxes we developed a Mathematica code, publicly available at \cite{SGREP_REPO}, together with tabulated values of $q^{-2}\dot{\bar{E}}^{\pm}_\tn{scal}$ as a function of $(r/M,a/M,\mu_s M)$. Further details on the implementation are given in Appendix \ref{appendix:implementation}. 

The energy emission drives the EMRI orbital evolution and, in the adiabatic approximation, the balance law between the binary binding energy and the GW flux $\dot{E} = - \dot{E}_\tn{GW}$ allows to compute the change in the orbital parameters, i.e. the radial and the azimuthal coordinates $(r,\phi)$. We set the initial phase $\phi_0$ to zero and the initial radius $r_0$  such that the EMRI evolves until the secondary reaches a plunging radius of $0.1M$ from the innermost stable circular orbit in $T=1$ year. 

\begin{comment}

\begin{equation}
\frac{dr}{dt}= \ - \dot{E}_\tn{GW} \frac{dr}{dE}\quad\ , \quad 
\frac{d\phi}{dt}= \Omega_\tn{p}\ ,  
\label{eq:orbital}
\end{equation}
with initial conditions $\phi_0=0$, and $r_0$ fixed such that the EMRI evolves until the secondary reaches a plunging radius of $0.1M$ from the innermost stable circular orbit in $T=1$ year.
\end{comment}

We model the emitted time-dependent gravitational waveform in the quadrupole approximation, finding the GW strain measured by the detectors $h(t,\vec{\theta})=F_+h_++F_\times h_\times$. This quantity depends on twelve parameters $\vec{\theta}=(\ln M,\ln m_p, \chi,d,\bar{\mu}_s,r_0,\phi_0,\theta_s,\phi_s,\theta_l,\phi_l,d_\tn{L})$, where $d_\tn{L}$ is the source luminosity distance, $\chi = a/M$ is the dimensionless spin parameter, and $\bar{\mu}_s = \mu_s M$. The LISA orbital motion is taken into account by the time-dependent pattern functions $F_{+,\times}$ which depend on the binary orientation $(\theta_s,\phi_s)$ and the spin direction $(\theta_l,\phi_l)$ in a solar barycentric frame (see\,\cite{Maselli:2021men} for further details on the waveform modeling and implementation).

Given two templates $h_{1,2}$ we define their inner product 
\begin{equation}
\langle h_1\vert h_2\rangle=4\Re\int_{f_\tn{min}}^{f_\tn{\rm max}}\frac{\tilde{h}_1(f)\tilde{h}^\star_2(f)}{S_n(f)}\dd f\ , \label{eq:sca_prod}
\end{equation}
where $\tilde{h}(f)$ is the Fourier transform of the time-domain signal, the $\star$ superscript identifies complex conjugation and $S_n$ is the LISA power spectral density, which includes the confusion noise of unresolved white-dwarf binaries \cite{Robson:2018ifk}. The signal-to-noise ratio (SNR) of a given waveform $h_1$ is then given by $\rho=\langle h_1\vert h_1\rangle^{1/2}$.
We also define the faithfulness between two templates 

\begin{equation}\label{eq:def_F}
\mathcal{F}[h_1,h_2]=\max_{\{t_c,\Phi_c\}}\frac{\langle h_1\vert 
	h_2\rangle}{\sqrt{\langle h_1\vert h_1\rangle\langle h_2\vert h_2\rangle}}\ ,
\end{equation}
with $(t_c,\Phi_c)$ time and phase offsets. This quantity provides an estimate of how much two waveforms differ, weighted by the detector sensitivity. We assume that for an SNR $\rho = 30$, two signals are distinguishable if $\mathcal{F} \lesssim \mathcal{F}_\tn{thr} =0.994$ \cite{Chatziioannou:2017tdw}.

In the limit of large SNR, the posterior distribution of $\vec{\theta}$ inferred by an EMRI detection can be approximated by a Gaussian centered around the true values $\vec{\hat{\theta}}$, with covariance ${\bf \Sigma}={\bf \Gamma}^{-1}$, where $\Gamma_{ij}=\langle \frac{\partial h}{\partial \theta_i}\vert\frac{\partial h}{\partial \theta_j}\rangle_{\vec{\theta}=\vec{\hat{\theta}}}$ is the Fisher information matrix, whose diagonal element $\sigma_i=\Sigma^{1/2}_{ii}$ corresponds to the statistical error of the $i$-th parameter, and $c_{\theta_i \theta_j}= \Sigma_{ij}/\sigma_{\theta_i}\sigma_{\theta_j}$ is the correlation coefficient between the parameters $\theta_i$, $\theta_j$. 
In this approach the SNR scales linearly with the inverse of the luminosity distance. Hereafter, we  scale $d_\tn{L}$ in order to have binaries with $\rho=150$, which is in the range of the expected SNRs of EMRI detections by LISA\,\cite{Babak:2017tow}.
Moreover, we fix $M=10^6M_\odot$, $\chi=0.9$, and $\theta_s=\phi_s=\pi/2$, $\theta_l=\phi_l=\pi/4$. 

\begin{figure}[htbp!]
    \centering
    \includegraphics[scale = 0.41]{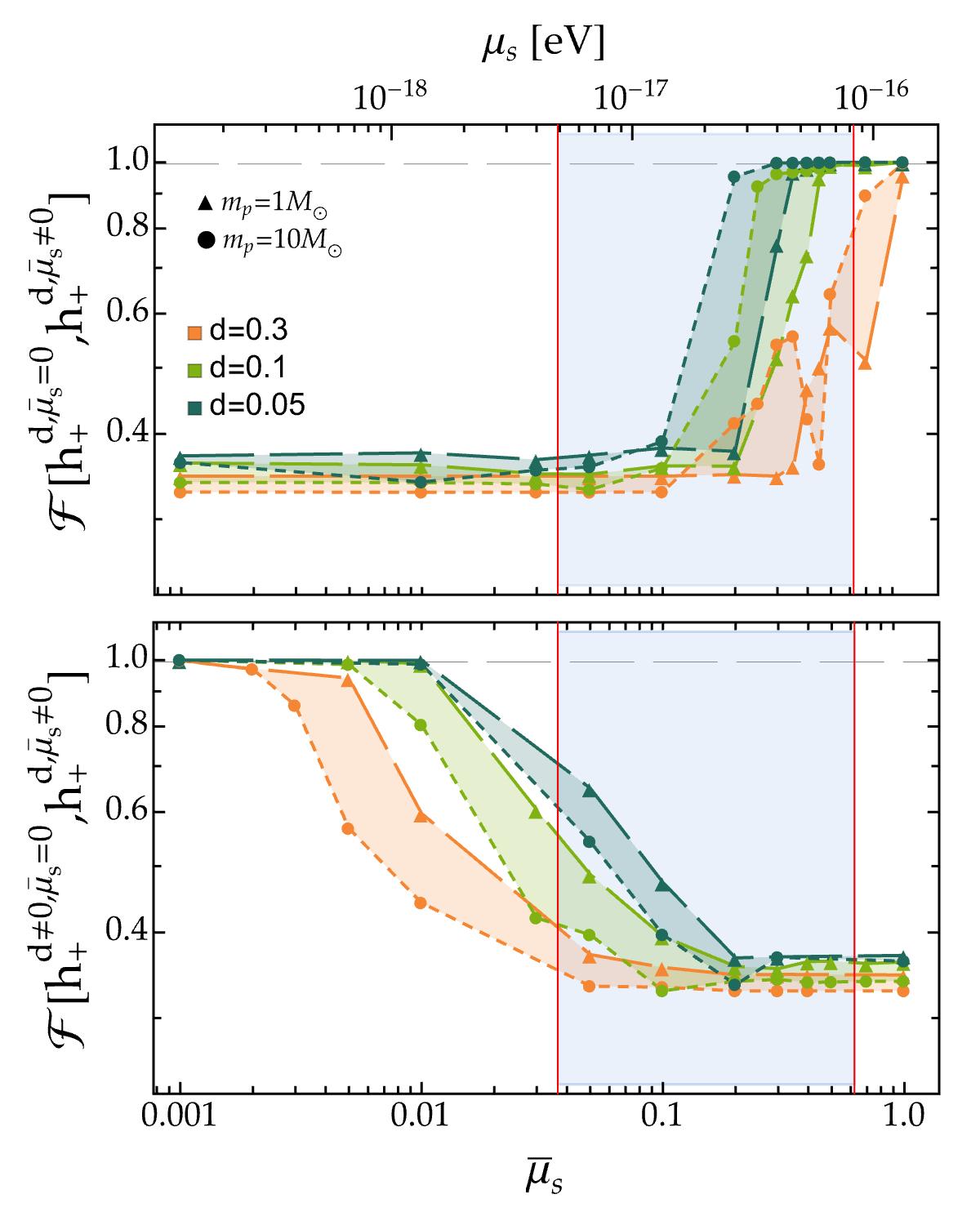}
    \caption{(Top) Faithfulness between a GW signal with `plus' polarisation 
    with $d=0$ and one with $d\neq0$, $\bar{\mu}_s\neq0$ for $12$ months of observation before the plunge. 
    We fix the primary mass and spin to $M = 10^6 M_{\odot}$  and $\chi=0.9$, respectively, while considering different values $m_p$ and $d$. The shaded region corresponds to  the range of scalar field masses which could be excluded by superradiance bounds (courtesy of R. Brito). (Bottom) Faithfulness  between  two signals with the same value of $d\neq0$, one having $\bar{\mu}_s=0$ and the other with $\bar{\mu}_s\neq0$. The horizontal dashed line corresponds to the threshold value $\mathcal{F}_{\tn{thr}}$. We consider the same EMRI configurations as in the top panel. 
}
    \label{fig:overlap_M6}
\end{figure}

\noindent{{\bf{\em Results.}}} 
We first study the distinguishability between the baseline GR model, i.e. assuming $(d,\bar{\mu}_s)=(0,0)$, and waveforms with non-vanishing values of the charge and of the scalar field mass. The top panel of Fig.~\ref{fig:overlap_M6} shows the faithfulness between the `plus' polarization $h_+$ computed in these two scenarios, for EMRIs with secondary mass of one and ten solar masses, as a function of $d$ and $\bar{\mu}_s$. 
As previously discussed, large values of $\bar{\mu}_s$ tend to suppress the GW flux at infinity, and hence the overall dissipative contribution of the scalar sector, as the energy emission at the horizon is subdominant.

Indeed, the faithfulness deteriorates rapidly as the scalar field mass decreases. For $0.05<d<0.1$, it lies below $\mathcal{F}_\tn{thr}$ for $\bar{\mu}_s \lesssim 0.3$ for the binaries we considered. Larger values of the scalar charge $(d=0.3)$ allow the two waveforms to be distinguishable for more massive scalar configurations, with $\bar{\mu}_s \gtrsim 0.7$. For a lighter secondary the faithfulness appears to reach $\mathcal{F}_\tn{thr}$ at a larger $\bar{\mu}_s$. However, the $d=0.3$ case is an outlier in this respect and also exhibits some additional peaks and troughs for larger values of  $\bar{\mu}_s$, which persists for larger values of $d$. The corresponding fluxes do not exhibit any remarkable difference from those corresponding to lower values of $d$ or $\bar{\mu}_s$, so it is not clear what causes these changes in the faithfulness for larger values of $d$ and $\bar{\mu}_s$.

We also note that for $\bar{\mu}_s\lesssim0.03$ ($\mu_s\lesssim 4\cdot10^{-18}$eV) the GR and the scalar waveforms are clearly distinguishable, with $\mathcal{F}_\tn{thr}\lesssim0.4$, regardless of the charge. 
Such estimates are complementary to other bounds which are expected to provide information on the existence of scalar fields in the gravity sector from future astrophysical probes. 
As an example, in both panels of Fig.~\ref{fig:overlap_M6} we draw as shaded regions the parameter space which can be potentially ruled out by superradiance constraints inferred from observations of massive BH binaries \cite{Brito:2015oca}. Our results suggest that, depending on $d$, EMRIs provide a new powerful channel to constrain both light and heavy fields, which do not fall within the superradiance window. 

As a step forward in this analysis we exploit the faithfulness to assess the minimum $\bar{\mu}_s$ which can be distinguished from the massless case. The bottom panel of Fig.~\ref{fig:overlap_M6} shows indeed the values of ${\cal F}$ computed between the gravitational waveform with `plus' polarization with either $\bar{\mu}_s=0$ or $\bar{\mu}_s\neq 0$, and fixed scalar charge. We consider the same binaries analysed in the top panel. 
Our results show that, for charges as small as $d\sim 0.05$, LISA could be able to distinguish fields with $\bar{\mu}_s\gtrsim 0.01$ ($\mu_s\sim 10^{-18}$eV) from their massless counterpart. This bound is larger by almost an order of magnitude if $d\gtrsim 0.3$.

\begin{figure*}[htbp!]
     \includegraphics[scale = 0.5]{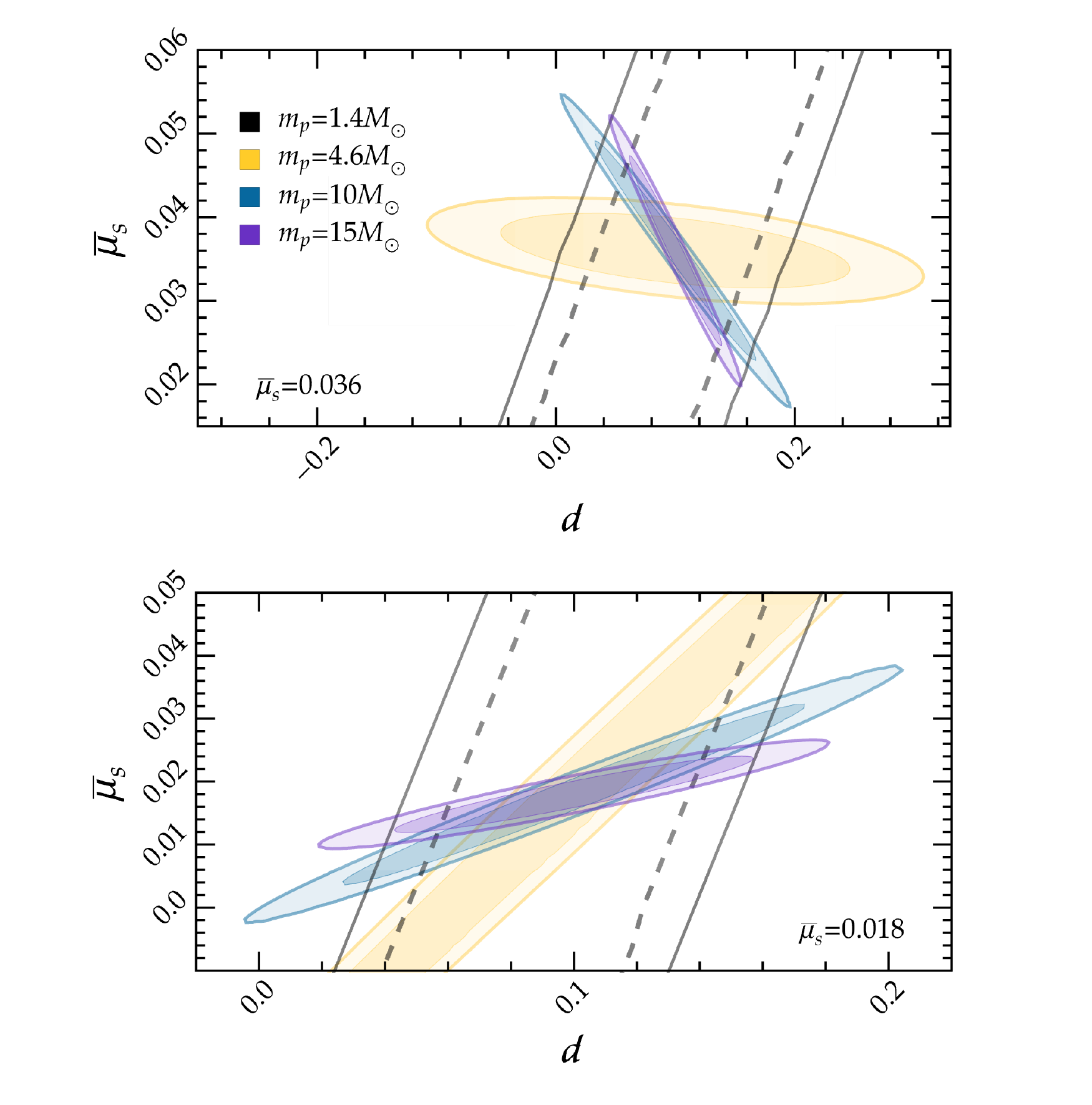}
    \includegraphics[scale = 0.5]{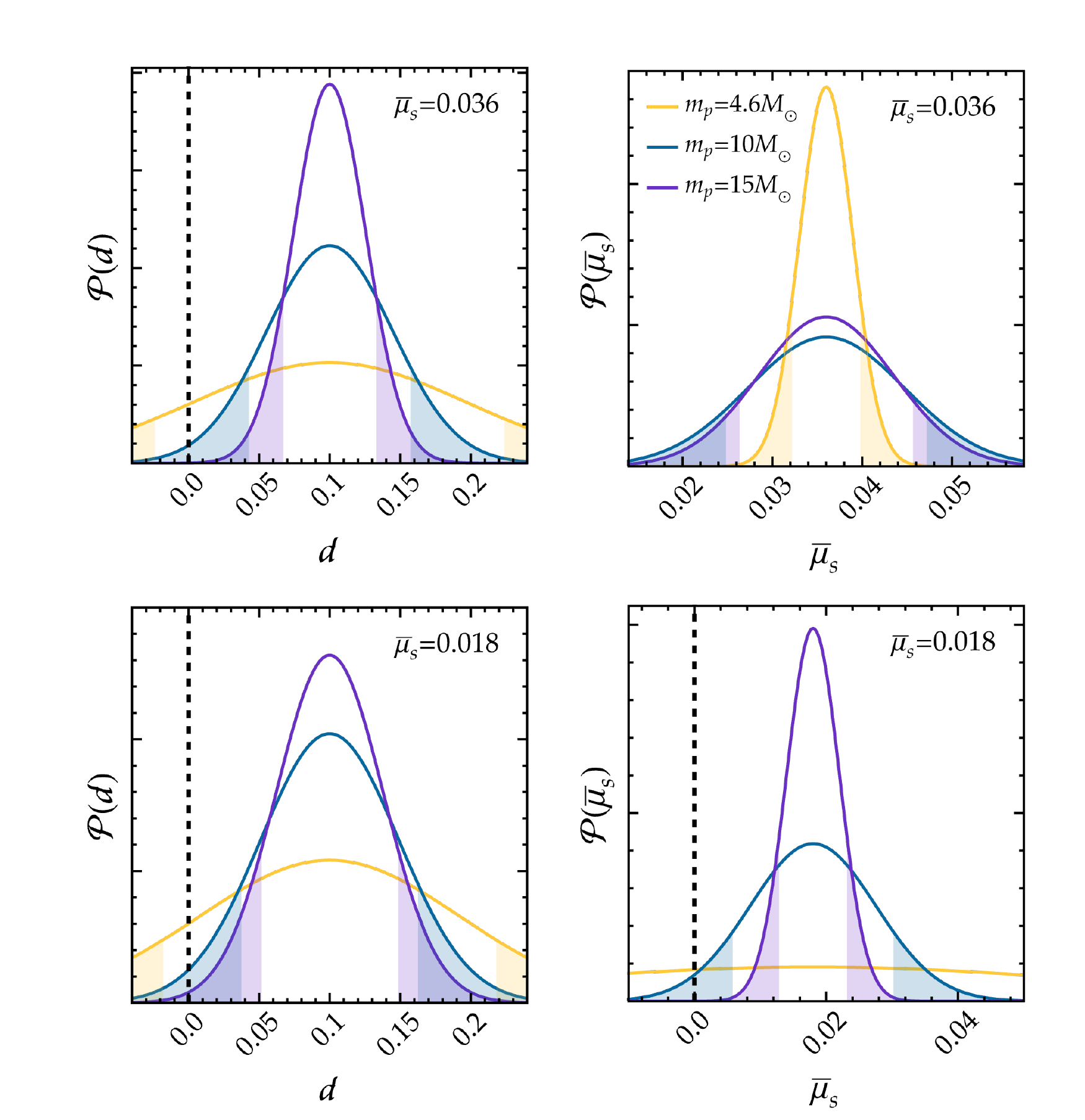}
    \caption{(Left column) Credible intervals at 68\% and 90\%   for the joint posterior distribution of the charge $d$  and the scalar field mass $ \bar{\mu}_s$. We consider EMRIs with $d=0.1$, $M=10^6 M_\odot$, $a=0.9M$,   different values of the secondary mass,  $ \bar{\mu}_s  = 0.018$ (top row),    and $ \bar{\mu}_s  = 0.036$ (bottom row). 
    (Right column) Marginal distributions for $d$    and $ \bar{\mu}_s$. The white area between shaded regions  provides 90\% of the probability distribution. The vertical lines identify the GR scenario with $d=\bar{\mu}_s=0$.}
    \label{fig:fisher}
\end{figure*}

The analysis developed so far however, takes only partially into account the correlations between the waveform parameters, which could hamper our ability to reconstruct the charge and the mass of the scalar field. The actual detectability of such parameters requires a more sophisticate analysis, based on the Fisher matrix approach. We apply the latter to LISA observations of prototype EMRIs with $d=0.1$, considering two values of $\bar{\mu}_s=(0.018,0.036)$, which lie outside the superradiance window highlighted in Fig.~\ref{fig:overlap_M6}, and for which the flux at infinity is significant throughout the entire inspiral.
The joint and marginal posterior distributions on $\bar{\mu}_s$ and $d$ derived for these systems are shown in the left and right columns of Fig.~\ref{fig:fisher}, respectively. A summary of the $1$-$\sigma$ uncertainties inferred for $\bar{\mu}_s$ and $d$ is reported in Table~\ref{tab:errors}, together with their correlation coefficients, which show how $\bar{\mu}_s$ and $d$ are strongly (anti-)correlated.

Errors on $d$ decrease as the mass ratio $m_p/M$ increases, for both values of $\bar{\mu}_s$. Binaries with $m_p\gtrsim 10M_\odot$ are able to exclude the $d=0$ case at more than $90\%$ credible level. For the EMRI configuration with $m_p=4.6M_\odot$, errors slightly deteriorate, with the null scenario ruled out at one sigma. Constraints on $\bar{\mu}_s$ show more variability. For the lowest injected value, $\bar{\mu}_s=0.018$, errors follow the same hierarchy observed for the scalar charge, with the measurement accuracy improving for heavier secondaries. In this setup however, $\bar{\mu}_s$ remains unconstrained for the EMRI with $m_p=4.6M_\odot$. This picture changes completely for the $\bar{\mu}_{s}=0.036$ case, in which the strongest bound is led by the lightest secondary. Binaries with $m_p=10M_\odot$ and $m_p=15M_\odot$ provide larger, and almost identical, errors.
Concerning the dependence of the results on the mass of the secondary, it is worth emphasising that we are considering one year of observation before the plunge, and hence secondaries with larger masses start as larger initial radii, where the scalar flux has a larger relative contribution. 

In comparison with the massless case, where the relative error on the scalar charge for the binary with $m_p=10M_\odot$ is $\simeq 4\%$ \cite{Maselli:2021men}, here it is larger: 
$\sigma_d/d \simeq 45\%$ and $49\%$ for $\bar{\mu}_s=0.036$ and $\bar{\mu}_s=0.018$ respectively. This is expected due to correlations with $\bar{\mu}_s$ which enters now as an additional parameter. Nevertheless, in all cases in which the probability distribution of $\bar{\mu}_s$ is constrained by the data, we are able to exclude the massless scenario at more than 90\% credible level.

\begin{table}[htbp!]
\centering
\begin{tabular}{ccccc}
\hline
$m_p [M_\odot]$ & $\bar{\mu}_s $ &$\sigma_{d}/d$ &  $\sigma_{\bar{\mu}_s}/\bar{\mu}_s$ & $c_{d\bar{\mu}_s}$    \\ 
\hline 
4.6&0.018 & 92\% & 243\% & 0.995\\
&0.036 & 97\% & 8\% & $-0.485$\\
\hline
10&0.018 & 49\% & 53\% & 0.984\\
&0.036 & 45\% & 24\% & $-0.990$\\
\hline
15&0.018 & 38\% & 22\% & 0.938\\
&0.036 & 26\% & 21\% & $-0.986$\\
\hline
\hline
\end{tabular}
\caption{1-$\sigma$ relative uncertainties and correlation coefficients on the charge and on the scalar field mass for the configurations shown in Fig.~\ref{fig:fisher}. We assume $d=0.1$ for all the binaries.}\label{tab:errors}
\end{table}

\noindent{{\bf{\em Discussion.}}}
Our results provide the first direct analysis on the capability of EMRI observations by the future space interferometer LISA to detect massive scalar fields and simultaneously measure the mass of the scalar and the scalar charge of the secondary. Our analysis assumes that the primary is adequately described by the Kerr spacetime at 
leading order in the mass ratio.
We have shown, using no-hair theorems \cite{Bekenstein:1995un,Sotiriou:2011dz} and EFT arguments \cite{Maselli:2020zgv}, that this is quite generically a valid assumption, provided that the primary is a black hole. 
Indeed our setup is theory-agnostic, and changes in the binary evolution
are uniquely determined by the scalar charge per unit mass of the EMRI secondary, $d$, and by the scalar field mass, $\mu_s$.
Therefore, ready-to-use templates for parameter estimation and phenomenological studies can be straightforwardly generated for a vast range of beyond-GR and beyond-Standard Model scenarios that contain a new massive scalar.

We have exploited such waveforms to assess the combined effect of nonvanishing $d$ and $\mu_s$. By computing the faithfulness between signals from uncharged and charged secondaries, we have shown that ultra-light scalar fields can leave a strong imprint on the GW emission, potentially detectable by LISA for a wide range of binary configurations. In particular, our results show how EMRIs provide a new observational window, complementary to other astrophysical probes, for detecting or constraining ultra-light scalar fields.

We have further investigated the actual constraints that LISA will be able to infer on the scalar charge and on field's mass, by performing a parameter estimation on prototype EMRI signals. Our results suggest that LISA will be able to measure, with a single event, {\it both} $d$ and $\mu_s$ accurately enough to potentially confirm the existence of an ultra-light scalar field at more than 90\% confidence level. 

Our analysis only focused on equatorial circular inspirals. Realistic scenarios EMRIs are expected to follow more complex trajectories along inclined and eccentric orbits. The effects of inclination are currently under investigation\,\cite{papeinclined}, while the inclusion of eccentricity  is expected to further enhance the distinguishability between signals with and without a scalar field \cite{Barsanti:2022ana, Zhang:2022rfr}. 
Improvements to our work would also include using fully-relativistic GW templates, performing a Bayesian analysis, and including post-adiabatic terms which take into account self-force corrections \cite{sperinew}. Considering the effects of resonances \cite{Yunes:2011aa} is also expected to further increase the distinguishability against GR signals, thus strengthening the results of our analysis.

\noindent{{\bf{\em Acknowledgments.}}}
This work makes use of the Black Hole Perturbation Toolkit. The authors would like to acknowledge networking support by the COST Action CA16104. T.P.S. acknowledges partial support from the STFC 
Consolidated Grant no. ST/T000732/1 and no. ST/V005596/1.
L.G. acknowledges financial support from the EU Horizon 
2020 Research and Innovation Programme under the 
Marie Sklodowska-Curie Grant Agreement 
no. 101007855.

\newpage

\section*{Appendix}

\subsection{Scalar perturbations}\label{appendix:scalar perturbations}
Hereafter we only discuss the scalar field component, since the tensor counterpart is well known in literature. 
Decomposing $\varphi$ in spheroidal harmonics allows to decouple the radial and the angular components (sum over the multipoles $\ell$ and $m$ is implicit):
\begin{equation}
    \varphi (t, r,\theta, \phi) = \int \dd\omega { \frac{\tilde{R}_{\ell m}(r,\omega)}{\sqrt{r^2+a^2}} S_{\ell m}(\theta,\omega) e^{i m \phi} e^{-i\omega t}}\ ,
    \label{psisdec}
\end{equation}
where $S_{\ell m}(\theta,\omega)$ are the spin-zero spheroidal functions. 
$\tilde{R}_{\ell m}$ satisfies the Schr\"oedinger-like equation

\begin{equation}
    \frac{d^2\tilde{R}_{\ell m}}{dr^2_\star}+V_s\tilde{R}_{\ell m}  = J_{\ell m} \ ,     \label{eq:master}
\end{equation}
where the tortoise coordinate $r_\star$ is such that $dr_\star/dr=(r^2+a^2)/\Delta$, with $\Delta=r^2+a^2-2Mr$. 
The effective potential $V_s$ is\,\cite{Teukolsky:1973ha}
\begin{align}
    V_s =\left[\omega - \frac{a m }{\rho^2}\right]^2 - \frac{\Delta}{\rho^8}\left[\lambda_{\ell m} \rho^4 + 2 M r^3 +\right. \nonumber\\ \left. a^2 (\Delta -2 M r)+\frac{\mu_s^2}{\rho^6}\right]\ , 
\end{align}
with $\rho^2 = r^2 + a^2 $, $\lambda_{\ell m} = \bar{\lambda}_{\ell m} +2 m a \sqrt{\omega^2-\mu_s^2} -2 m a \omega $, and $\bar{\lambda}_{\ell m}$ being the angular eigenvalue associated to $S_{\ell m}(\theta,\omega)$.
The source term, for a particle moving on a circular orbit with radius $r_\tn{p}$ and (prograde) angular frequency $\Omega_p=M^{1/2}/(r_p^{3/2}+ a M^{1/2})$, is
\begin{equation}
   J_{\ell m } =  -  d \frac{4 \pi m_\textnormal{p} \Delta}{\sqrt{a^2+r^2}}  \frac{S_{\ell m}^{\star}\left(\pi/2\right)}{u^t} \delta  (r - r_\textnormal{p}) \delta (\omega - m \Omega_\textnormal{p})\,,  
   \label{eq:source}
\end{equation}
where $u^t$ is the time component of the particle four-velocity.

To solve Eq.\,\eqref{eq:master} for $\tilde{R}_{\ell m}$,
we first find the solution of the homogeneous problem, $\tilde{R}^{\mp}_{\ell m}$ which satisfies the condition of purely ingoing/outgoing wave at the horizon/infinity (see App.\,\ref{appendix:implementation}). 

The general solution $\tilde{R}_{\ell m}$ is then obtained by integrating the 
former over $J_{\ell m}$.
 
The energy fluxes of the scalar field at the horizon and at infinity are then found in terms of the asymptotic values of $\tilde{R}_{\ell m}$:
\begin{equation}
    \dot{E}^{\mp}_\textnormal{scal} = \frac{1}{16 \pi} \sum_{\ell =1}^\infty\sum_{m=-\ell}^\ell\omega k_{\mp} |Z^{\mp}_{\ell m}|^2\ ,\label{math:scalflux}
\end{equation}
with $\omega = m \Omega_\textnormal{p}$, $k_+ = \sqrt{\omega^2-\mu_s^2}$, $k_-=\omega - m \Omega_h$, $\Omega_h = a/(2Mr_h)$, $r_h = M + \sqrt{M^2 - a^2}$ and 
\begin{equation}
Z^{\mp}_{\ell m} = 
\tilde{R}^{\mp}_{\ell m} (r_{\star}\rightarrow\mp \infty) 
\int^{+\infty}_{-\infty} \frac{\tilde{R}^{\pm}_{\ell m} J_{\ell m} \dd r_\star}{W}\ ,
\end{equation}
where $W= \tilde{R}_{\ell m}^{'+} \tilde{R}_{\ell m}^- - \tilde{R}_{\ell m}^+ \tilde{R}_{\ell m}^{'-}$ is the Wronskian and primes denote derivatives with respect to $r_{\star}$. 

Since $\tilde{R}^{+}_{\ell m} (r_{\star}\rightarrow+ \infty)=e^{-i\omega r_*}\Theta(m \Omega_\textnormal{p} -\mu_s)$ with $\Theta$ Heaviside function, 
\begin{equation}
|Z^{+}_{\ell m}|^2 = \left|\int^{+\infty}_{-\infty} \frac{\tilde{R}^{-}_{\ell m} J_{\ell m} \dd r_\star}{W} \right|^2 \Theta (m \Omega_\textnormal{p} - \mu_s)\,.
\end{equation}

\subsection{Implementation}\label{appendix:implementation}
We have computed the gravitational and the  massless scalar fluxes using the  Black Hole Perturbation Toolkit \cite{BHPToolkit}, while for the massive scalar fluxes we have developed a Mathematica code, publicly available at \cite{SGREP_REPO}. To solve the scalar equation \eqref{eq:scal} we have adopted a standard Green function approach, in which we first compute the associated homogeneous solutions satisfying the following boundary conditions: 
\begin{align}
\tilde{R}_H(r)&=\sum^{n_h}_{n=0} e^{- i  k_- r_\star(r)}a_n(r-r_h)^n \ ,\\
\tilde{R}_\infty(r)&=\sum^{n_\infty}_{n=0} e^{i k_+   r_\star(r)} r^{\frac{i \mu_s^2 M} {\sqrt{\omega^2-\mu_s^2}}}   \frac{b_n}{r^n} \ ,
\end{align}
corresponding to purely ingoing and purely outgoing solutions at the horizon $r_h$ and at spatial infinity, respectively. The coefficients $(a_n,b_n)$ are obtained by solving the homogeneous equation at each order in ($r - r_h$) and $1/r$, with 
$a_0=b_0=1$. The order of the expansion in our code is set to $n_h=n_\infty=n_\tn{max}=4$. This choice guarantees very accurate boundary conditions for the numerical integration. Indeed, the relative difference in the dominant $\ell=m=1$ mode of the scalar flux computed with $n_\tn{max}=3$ and $n_\tn{max}=4$ is $\lesssim 10^{-10} \%$ within the integration domain $2.4M\leq r \leq 15M$ and $0.01\leq \bar{\mu}_s  \leq 1$.

We integrate the field's equations from $r_1 = r_h + 2\epsilon$, with $\epsilon = 10^{-5}$, to a value at infinity, $r_2$, which is either fixed to (i) $r_2=1000/\text{Abs}[\omega] $ if  $\omega > \mu_s$ or (ii) such that $\tilde{R}_\infty (r_2) \sim 10^{-50}$ if $\omega < \mu_s$. The last condition is chosen to avoid accuracy problems due to the exponential decay of $\tilde{R}_\infty (r_2)$ for wavelengths smaller than the scalar field Compton mass. 

The total gravitational and scalar emissions have been computed by summing over the modes $(\ell,m)$, with $\ell_\text{min}\leq \ell \leq \ell_\text{max}$ and  $-\ell \leq m \leq \ell$. For the gravitational (scalar) perturbations, $\ell_\text{min}=2$ $(\ell_\text{min}=1)$. 
For the faithfulness calculations we choose $\ell_{\text{max}}=10$ both for the gravitational and the scalar computations. For the gravitational emission, the relative difference between the fluxes with $\ell_{\text{max}}=9$ and with $\ell_{\text{max}}=10$ is $\sim 0.1 \%$ for $r=2.5 M$ and $\sim 10^{-4}\, \%$ for $r=8M$. For the scalar emission, the relative differences for different values of the scalar field mass are reported in Table \ref{tab:interpolation}. We observe that the relative difference is $<1\%$ for $r=2.5M$ and $<10^{-2}\,\%$ for $r=8M$, with scalar field mass in the range $0.001 \leq  \bar{\mu}_s  \leq 0.1$. For larger values of the scalar mass, $ \bar{\mu}_s \sim 1 $, the relative difference for $r=2.5M$ is larger.  

Finally, in order to derive the EMRI phase evolution, we interpolate the  GW fluxes over the radial coordinate using a built-in function of Mathematica. For a fixed primary spin and scalar field mass, we compute the fluxes over a grid of 251 points uniformly spaced in $u = (r-0.9 \ r_\text{ISCO})^{1/2}$ within $[u(r_\text{min}),u(r_\text{max})]$, where $r_ \text{max}= r_\text{min}+13M$ and $r_\text{min} = r_\text{ISCO} + 0.1M$, with $r_\text{ISCO}$ being the value of the radial coordinate at the ISCO as a function of the primary spin $a$. 

In order to have an estimate of the error brought by the interpolation, we  
computed the scalar fluxes for some orbital radii which don't lie on the grid defined above, and compared them with those obtained through the grid interpolation. 
The results are given in Table \ref{tab:interpolation}. The relative difference  
is $\lesssim 10^{-3}$ for $ \bar{\mu}_s  \lesssim0.1$. For higher values of the scalar field mass, the relative difference increases, since the flux at infinity vanishes for the modes with lower $m$ and the total flux experiences large variations between close points on the grid.

\begin{table*}[htbp!]
\centering
\begin{tabular}{ccccccc}
\hline
$ \bar{\mu}_s $ &   $r/M$ &  $\dot{E}^{\ell_{\text{max}}=10}_\text{scal}$ &$\dot{E}^\text{{int}}_\text{scal}$ & Rel. Diff. & $\dot{E}^{\ell_{\text{max}}=9}_\text{scal}$ & Rel. Diff. \\
  \hline
0.001 &2.5  &$7.562\times10^{-4}$ &  $7.562\times10^{-4}$ & $<10^{-8} \ \%$ &$7.528\times10^{-4}$ &  $<1\%$\\
    &8 &$1.542\times10^{-5}$ &$1.542\times10^{-5}$  & $<10^{-3} \ \%$ & $1.542\times10^{-5}$ & $<10^{-3} \%$\\ 
\hline
0.01 &2.5  &$7.550\times10^{-4}$   & $7.550\times10^{-4}$& $<10^{-8} \ \%$ & $7.515\times10^{-4}$ & $<1 \%$ \\
    &8 &$1.440\times10^{-5}$ &$1.440\times10^{-5}$ &  $<10^{-3} \ \%$&$1.440\times10^{-5}$ &$<10^{-3} \%$ \\ 
\hline    
0.1 &2.5  &$6.399\times10^{-4}$ &$6.399\times10^{-4}$  & $<10^{-8} \ \%$ &$6.365\times10^{-4}$& $<1 \%$\\
    &8 &  $-4.044\times10^{-7}$  &$-2.809\times10^{-7}$ & $\simeq 30\%$ &  $-4.044\times10^{-7}$ & $<10^{-2} \%$\\ 
\hline
1 &2.5   &$-3.146\times10^{-6}$&$-1.195\times10^{-6}$  &  $\simeq 60\%$ & $-3.541\times10^{-6}$ &   $\simeq 10\%$\\
    &8 & $-4.984\times10^{-13}$ & $-3.620\times10^{-13}$ & $\simeq 30\%$ & $-4.984\times10^{-13}$ & $<10^{-7} \ \%$\\ 
\hline
\hline
\end{tabular}
\caption{Relative difference between 
scalar energy flux computed up to $\ell=10$ and i) interpolated values, ii) values computed up to $\ell = 9$. The primary spin is $a/M=0.9$. The superscript \text{``int"} identifies the interpolated fluxes. }\label{tab:interpolation}
\end{table*}

\begin{figure}[htbp!]
    \centering
    \includegraphics[scale = 0.5]{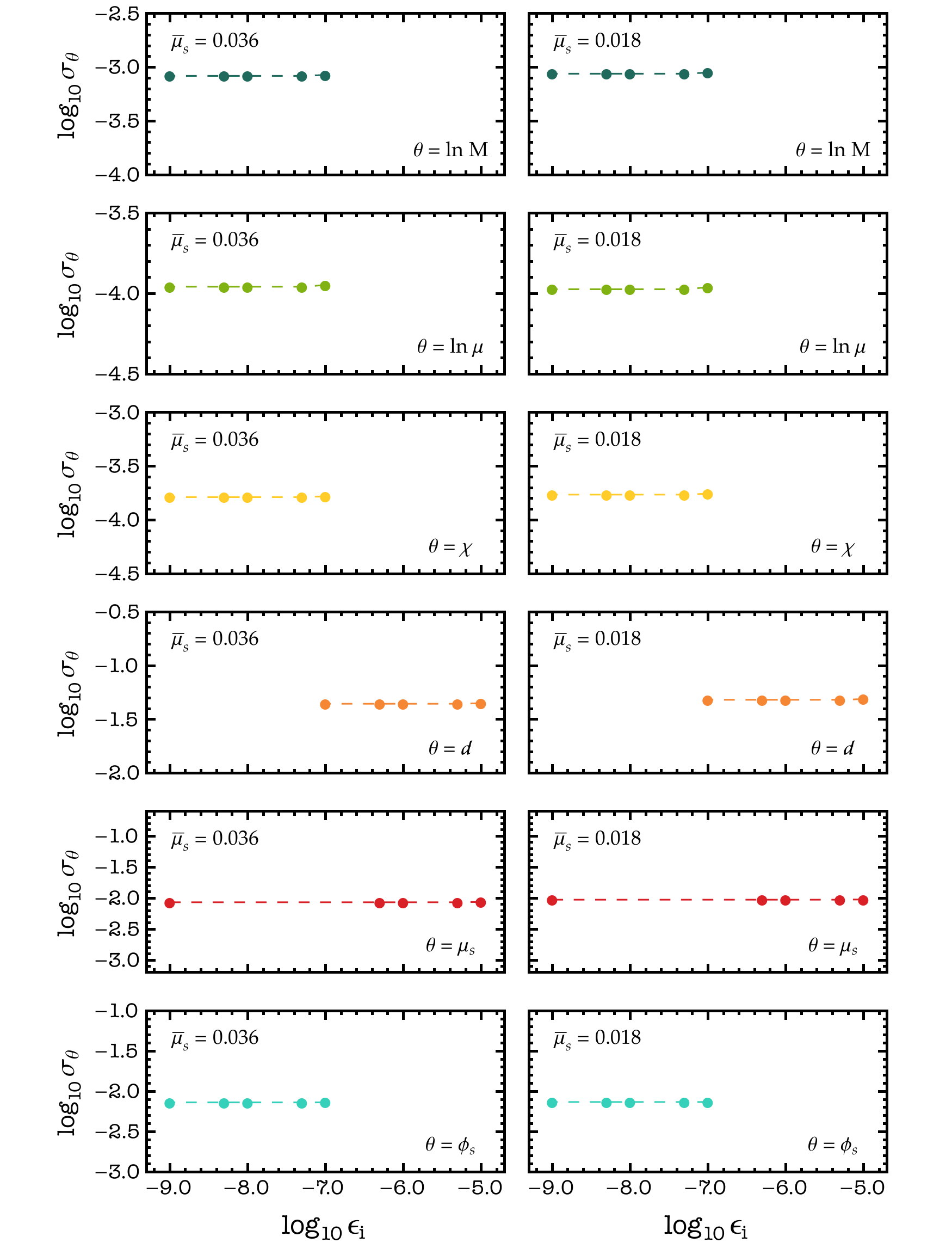}
    \caption{Errors on the binary parameter as a function of the shift. Injected parameters are $M=10^6 M_\odot$, $m_p=10M_\odot$, $\chi=0.9$, $d=0.1$, and  $\bar{\mu}_s=0.036$, $0.018$ for the left and right 
    column, respectively.}\label{fig:stability}
\end{figure}

\begin{figure}[htbp!]
    \centering
    \includegraphics[scale = 0.6]{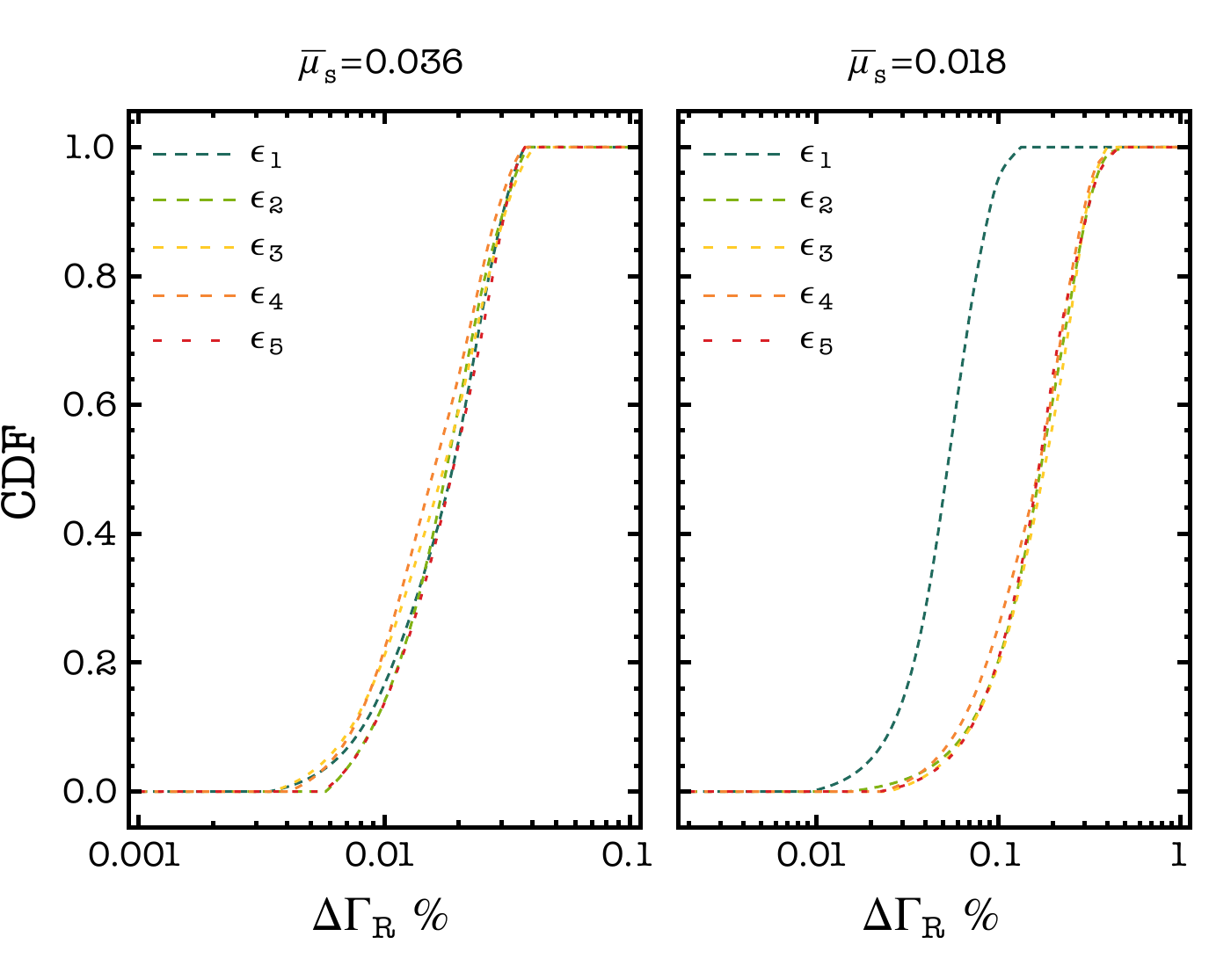}
    \caption{Cumulative distribution for the maximum relative error between unperturbed and perturbed Fisher matrices with elements shifted by random numbers drawn from a uniform distribution. 
    Curves with various colors refer to Fisher matrices computed with a different choice of the numerical derivative shifts, also shown in Fig.~\ref{fig:stability}. 
    We consider EMRI with $M=10^6 M_\odot$, $m_p=10M_\odot$, $\chi=0.9$, $d=0.1$, and $\bar{\mu}_s=0.036$, $0.018$ for the left and right column, respectively.}\label{fig:stability2}
\end{figure}

Unlike the faithfulness analysis, errors computed through the Fisher matrix approach have been derived by considering only the fundamental mode of the scalar flux, i.e. assuming $\ell_\tn{max} = 1$ in the sum \eqref{math:scalflux}.  This choice is dictated by the computational cost needed to invert the Fisher matrix, whose stability requires very accurate fluxes,  
computed on the grid $(r/M, \chi,  \bar{\mu}_s)$ with 205 digits of input precision (we refer the reader to Ref.\,\cite{Maselli:2021men} for an extensive discussion on the stability requirements of the Fisher matrix). Derivatives of the template with respect to the binary parameters are fully numerical, obtained by using a 11-points stencil. 
To compute finite differences we have sampled the spin parameter between $0.89\leq \chi \leq 0.91$ in steps of $\Delta \chi = 0.01$, and the scalar field mass in an interval of 11, equally spaced points, in steps of $\Delta  \bar{\mu}_s = 0.002$ (centered around the two injected values we considered, i.e. $ \bar{\mu}_s=0.018$ and $ \bar{\mu}_s=0.036$). 
Note that finite differences require a careful choice of a step-size $\epsilon$, which controls the shift for each of the twelve parameters of the waveform. We have varied these coefficients within a broad range of values to assess the stability of our Fisher matrices. A summary of this analysis is displayed in Fig.~\ref{fig:stability}, which shows the errors on some parameters as functions of $\epsilon$. To further assess the overall stability of the Fisher matrices we have studied how the errors on the parameters change under small perturbations. 
To this aim we have built a matrix $\textbf{R}$ with entries randomly drawn from a uniform distribution $U\in [-10^{-3},10^{-3}]$. We have then computed the inverse $({\bf \Gamma} + \textbf{R})^{-1}$ and the maximum relative error with respect to the unperturbed configuration, $\Delta{\bf  \Gamma}_\textbf{R} = {\rm max}[({\bf  \Gamma} + \textbf{R})^{-1}/{[\bf \Gamma}^{-1} - \textbf{I}]$.
We have iterated this procedure 100 times, to build up statistics for the maximum error. The cumulative distribution of $\Delta{\bf  \Gamma}_\textbf{R}$ for some of the EMRI configurations we considered is shown in Fig.~\ref{fig:stability2}, proving that our calculations are extremely stable with more than 90\% of the population having $\Delta{\bf  \Gamma}_\textbf{R}\lesssim 0.1\%$. Similar results hold for all the binaries we focused on.

\begin{figure}[htbp!]
    \centering
    \includegraphics[scale = 0.52]{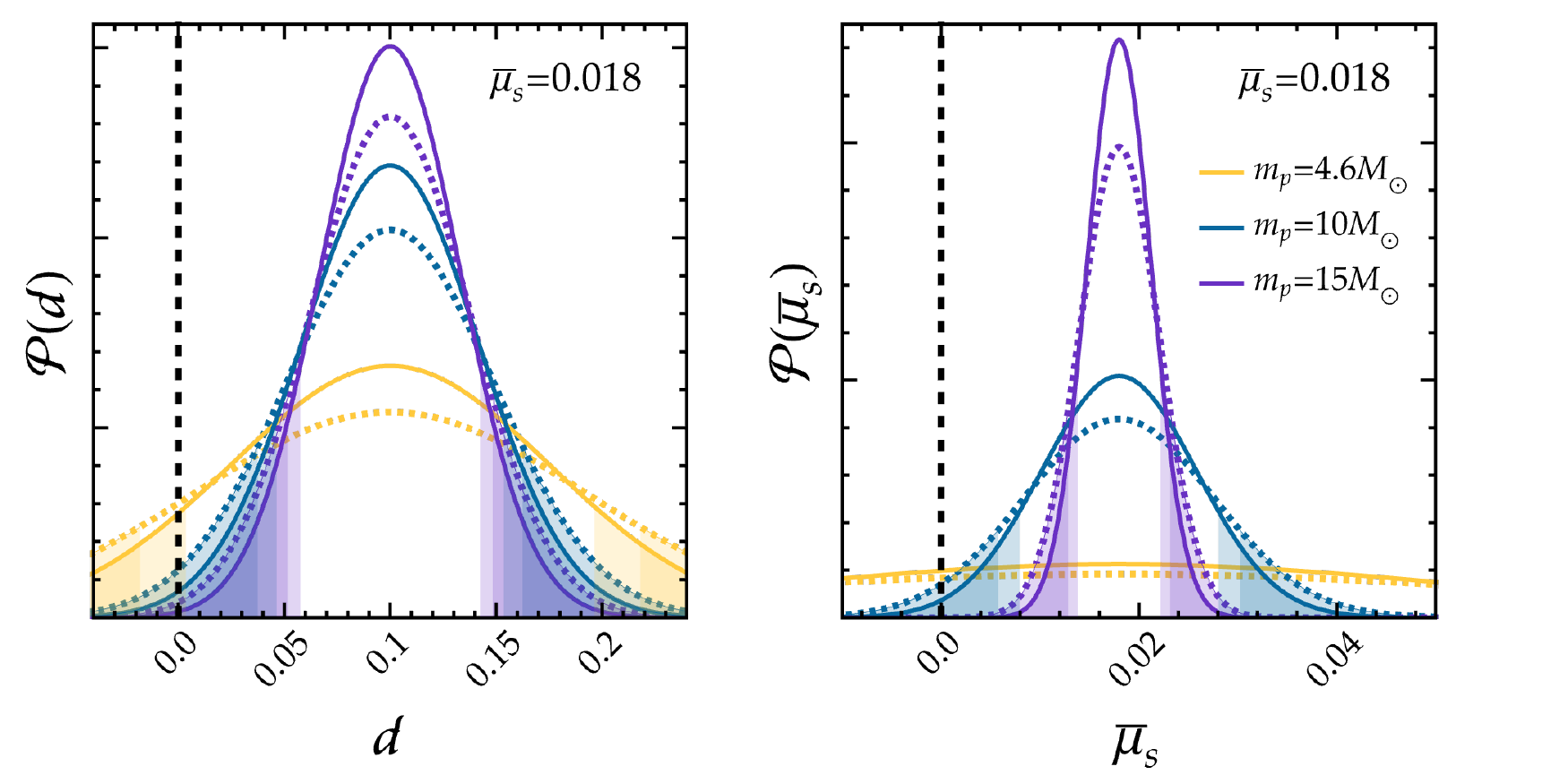}
    \caption{Marginal distributions for $d$ (left plot) and $\mu_s$ (right plot). Dashed and continuous curves are relative to inspirals in which the scalar flux takes into account only     the $\ell_\text{max}=1$ or up to $\ell_\text{max}=2$, respectively. The vertical lines identify the GR scenario with $d=\bar{\mu}_s=0$.}\label{fig:fihser_lm22}
\end{figure}

Finally, to quantify the bias that could arise in the error analysis by neglecting multipoles larger than $\ell=1$ for the scalar flux, we have performed a Fisher analysis for a configuration with $ \bar{\mu}_s  = 0.018$, by also including the $\ell=2$ and the $\ell =3$ modes within the sum \eqref{math:scalflux}. The probability distributions for the charge and for the scalar field mass obtained by adding the $\ell=2$ component are presented in Fig. \ref{fig:fihser_lm22}. 
Dashed and solid curves correspond to assuming either  $\ell_\text{max}=1$, or $\ell_\text{max}=2$. As expected, constraints on both $ \bar{\mu}_s$ and $d$ improve by adding the quadrupole contribution, making our choice rather conservative in terms of the constraints that LISA would be able to infer from EMRI observations. Table \ref{tab:errorsl} shows a comparison of the relative errors obtained by also adding the $\ell=3$ mode. 
For the binary configurations we considered, our results suggest that the uncertainties 
on both the parameters tend to saturate already with the inclusion of the third multipole.

\begin{table*}[htbp!]
\centering
\begin{tabular}{ccccccc}
\hline
$m_p [M_\odot]$ &$\sigma^{\ell_\text{max}=1}_{d}/d$ & $\sigma^{\ell_\text{max}=2}_{d}/d$ & $\sigma^{\ell_\text{max}=3}_{d}/d$ & $\sigma^{\ell_\text{max}=1}_{ \bar{\mu}_s}/ \bar{\mu}_s$ & $\sigma^{\ell_\text{max}=2}_{ \bar{\mu}_s}/ \bar{\mu}_s$ &  $\sigma^{\ell_\text{max}=3}_{ \bar{\mu}_s}/ \bar{\mu}_s$  \\ 
\hline 
4.6 & 92\% & 75\% & 78\%&243\% & 198\% &190\%\\
10 & 49\% & 42\% & 44\%&53\% & 44\% & 41\%\\
15 & 38\% & 33\% & 35\%&22\% & 18\% & 17\%\\
\hline
\hline
\end{tabular}
\caption{Relative percentage errors on the parameters $(d, \bar{\mu}_s)$ for $m_p = (4.6,10,15)M_\odot$ assuming different $\ell_\tn{max}$ in the scalar energy flux sum \eqref{math:scalflux}. The injected values are $d=0.1$ and $ \bar{\mu}_s=0.018$. }\label{tab:errorsl}
\end{table*}

\subsection{Fluxes and dephasing}\label{app:fluxes}
\begin{figure}[htbp!]
    \centering
    \includegraphics[scale = 0.5]{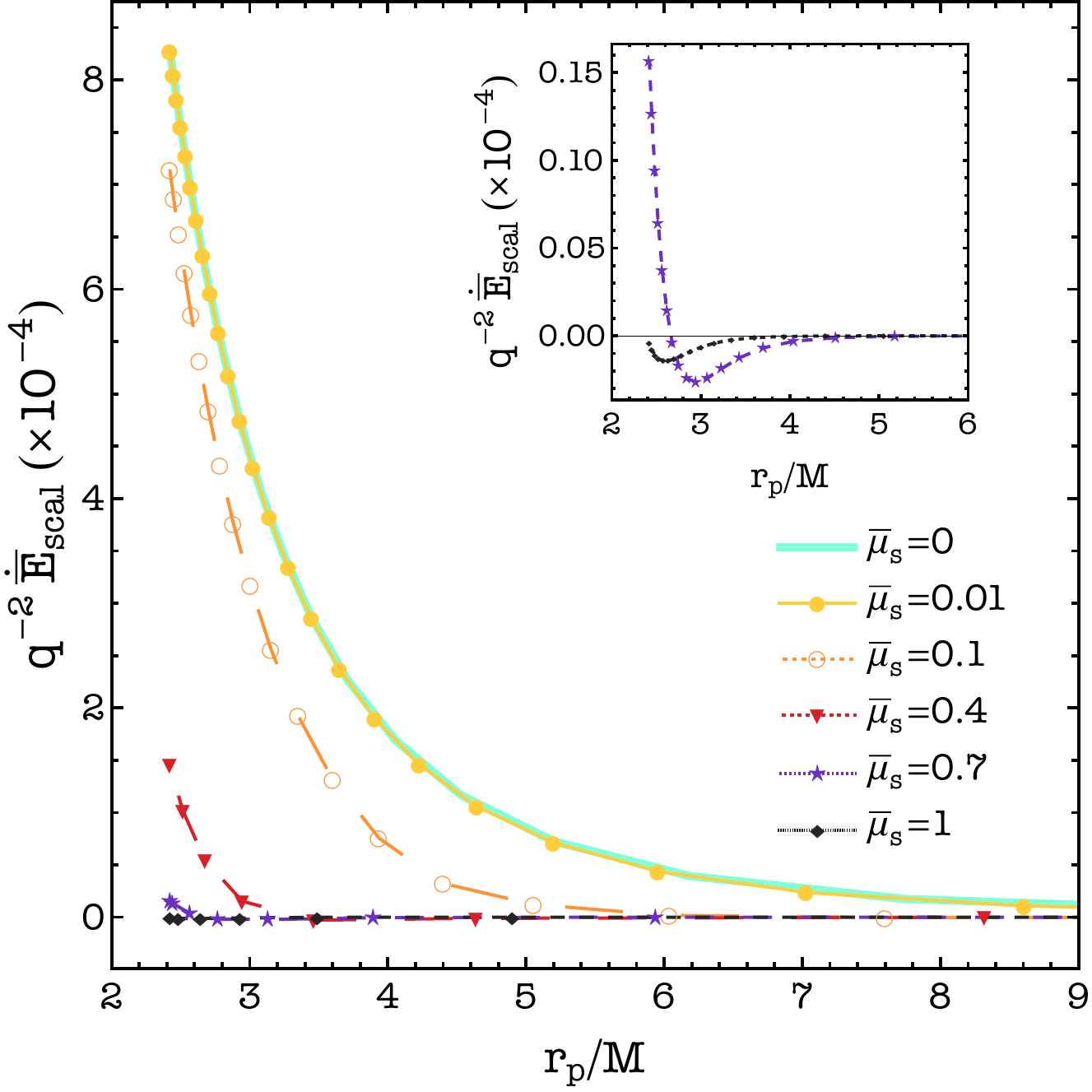}
    \caption{Total massive scalar flux up  to $\ell = 10$ as a function of secondary    orbital radius, for different values of the scalar field mass. Resonances are excluded. 
    The inset shows the fluxes for $ \bar{\mu}_s  = (0.7,1)$.     The primary spin is fixed to $\chi=0.9$. The continuous line refers to the massless case.}
    \label{fig:flux}
\end{figure}

In this Appendix we discuss further details on the EMRI scalar emission, and how it affects the binary orbital evolution. Fig.\,\ref{fig:flux} shows the behaviour of $\dot{\bar{E}}_\tn{scal}$ as a function of the secondary orbital radius $r_p/M$, for different values of the scalar field mass $ \bar{\mu}_s$, for a BH with dimensionless spin parameter $\chi=0.9$. 
The flux contains multipoles up to $\ell_\tn{max} = 10$. For $\bar{\mu}_s  \lesssim 0.1$ the qualitative behaviour of the energy emission is the same as in the massless case, with a maximum relative discrepancy of $\sim 10\%$ within the orbital range we considered. By increasing $\bar{\mu}_s$, the suppression of the flux at infinity becomes more and more relevant, quenching the overall emission. Note that since $\dot{\bar{E}}_\tn{scal}\sim\Theta(m \Omega_p-\mu_s)$, for larger values of $\bar{\mu}_s$ such cancellation affects orbits closer to the ISCO. This feature appears more evident within the inset of Fig.~\ref{fig:flux}, which shows the scalar flux for the most massive cases, $\bar{\mu}_s=0.7$ and $\bar{\mu}_s=1$.

The presence of the scalar emission modifies the evolution of the radial and orbital phase. The latter in particular can be used to provide a first quantitative assessment of the change induced by the scalar field to the EMRI dynamics. To this aim we compute the dephasing, i.e. the phase difference integrated over a given period of observation, between an inspiral with $d,\bar{\mu}_s\neq 0$ and one with $d,\bar{\mu}_s=0$, 
\begin{equation}
    \Delta \phi= 2 \int^{T_\tn{obs}}_{0} [ \Omega_{d,\bar{\mu}_s = 0} - \Omega_{d,\bar{\mu}_s \neq 0} ]\  dt\ . 
\label{eq:deph}
\end{equation}
For EMRIs observed by LISA with a SNR of $\rho = 30$, we can introduce a threshold of $\vert\Delta \phi\vert = 0.1$ radians, beyond which the two signals can potentially be distinguished \cite{Bonga:2019ycj}.

\begin{figure}[htbp!]
    \centering
    \includegraphics[scale = 0.55]{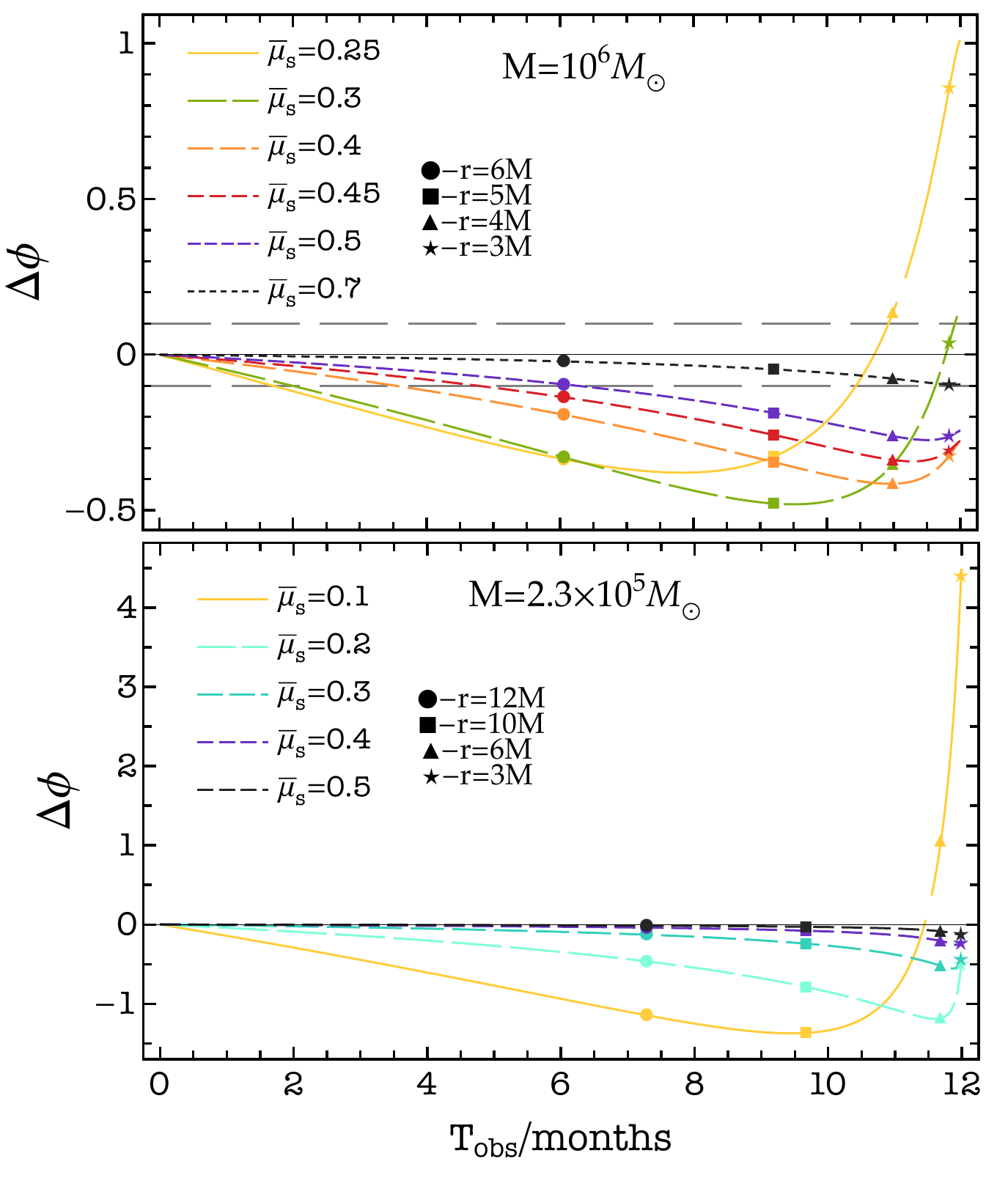}
    \caption{(Top) Dephasing accumulated in one year before the plunge, plotted as a function of the time of observation for a binary system   with $M=10^6M_\odot$, $\chi=0.9$ and $m_p=10M_\odot$,   for different values of the scalar field mass. The difference in the phase evolution is computed between an inspiral with $d=0$ and one with $d=0.1$. (Bottom) 
    Same as top panel but for a primary mass of  $M=2.3\times10^6M_\odot$. Different values of the   scalar field mass are considered.}
    \label{fig:dephasing}
\end{figure}

Figure~\ref{fig:dephasing} shows the dephasing as a function of the time of observation, for different values of the scalar field mass, and for two primary configurations, $M=10^6 M_\odot$ (top panel) and $M=2.3 \times 10^5 M_\odot$ (bottom panel), with both $\chi=0.9$. 
The markers identify the time after which the secondary with $(d,\bar{\mu}_s)=0$ crosses a particular radius, shown in the legend of the plots. For the charged configurations we fix $d=0.1$. The dashed grey horizontal lines identify the LISA phase resolution threshold at $\Delta \phi= \pm 0.1$ radians.

For each $\bar{\mu}_s$, regardless of $M$, $\Delta \phi$ decreases during the evolution before increasing in the last few months of the inspiral. This behavior is due to the  presence of the scalar flux at infinity, which shifts the total scalar emission from negative to positive values, thus accelerating the inspirals with $(d,\bar{\mu}_s)\neq 0$. 
For $M=10^6 M_\odot$ the crossing between negative and positive values of the dephasing occurs for $\bar{\mu}_s \lesssim 0.3$ ($\mu_s \lesssim 4 \times 10^{-17} \tn{eV}$), while for $M=2.3 \times 10^5 M_\odot$ it happens for $\bar{\mu}_s \lesssim 0.1$ ($\mu_s \lesssim 6 \times 10^{-17} \tn{eV}$). This analysis shows that light fields, with $\bar{\mu}_s \ll 1$, lead in general to differences in the phase evolution 
which in modulo are larger than the detectability threshold, thus potentially measurable by LISA. The actual constraints on the scalar field properties require though to fully take into account correlations between the waveform parameters, as discussed in the Fisher matrix analysis presented in the main text.

\bibliographystyle{utphys}
\bibliography{Ref}

\providecommand{\href}[2]{#2}\begingroup\raggedright\begin{thebibliography}{10}

\bibitem{LIGOScientific:2021djp}
{\bfseries LIGO Scientific, VIRGO, KAGRA} Collaboration, R.~Abbott {\em
  et~al.}, ``{GWTC-3: Compact Binary Coalescences Observed by LIGO and Virgo
  During the Second Part of the Third Observing Run},''
  \href{http://arxiv.org/abs/2111.03606}{{\ttfamily arXiv:2111.03606 [gr-qc]}}.

\bibitem{Audley:2017drz}
{\bfseries LISA} Collaboration, P.~Amaro-Seoane {\em et~al.}, ``{Laser
  Interferometer Space Antenna},''
  \href{http://arxiv.org/abs/1702.00786}{{\ttfamily arXiv:1702.00786
  [astro-ph.IM]}}.

\bibitem{Seoane:2021kkk}
P.~A. Seoane {\em et~al.}, ``{The effect of mission duration on LISA science
  objectives},'' \href{http://dx.doi.org/10.1007/s10714-021-02889-x}{{\em Gen.
  Rel. Grav.} {\bfseries 54} no.~1, (2022) 3},
  \href{http://arxiv.org/abs/2107.09665}{{\ttfamily arXiv:2107.09665
  [astro-ph.IM]}}.

\bibitem{Laghi:2021pqk}
D.~Laghi, N.~Tamanini, W.~Del~Pozzo, A.~Sesana, J.~Gair, S.~Babak, and
  D.~Izquierdo-Villalba, ``{Gravitational-wave cosmology with extreme
  mass-ratio inspirals},'' \href{http://dx.doi.org/10.1093/mnras/stab2741}{{\em
  Mon. Not. Roy. Astron. Soc.} {\bfseries 508} no.~3, (2021) 4512--4531},
  \href{http://arxiv.org/abs/2102.01708}{{\ttfamily arXiv:2102.01708
  [astro-ph.CO]}}.

\bibitem{Berry:2019wgg}
C.~P.~L. Berry, S.~A. Hughes, C.~F. Sopuerta, A.~J.~K. Chua, A.~Heffernan,
  K.~Holley-Bockelmann, D.~P. Mihaylov, M.~C. Miller, and A.~Sesana, ``{The
  unique potential of extreme mass-ratio inspirals for gravitational-wave
  astronomy},'' \href{http://arxiv.org/abs/1903.03686}{{\ttfamily
  arXiv:1903.03686 [astro-ph.HE]}}.

\bibitem{McGee:2018qwb}
S.~McGee, A.~Sesana, and A.~Vecchio, ``{Linking gravitational waves and X-ray
  phenomena with joint LISA and Athena observations},''
  \href{http://dx.doi.org/10.1038/s41550-019-0969-7}{{\em Nature Astron.}
  {\bfseries 4} no.~1, (2020) 26--31},
  \href{http://arxiv.org/abs/1811.00050}{{\ttfamily arXiv:1811.00050
  [astro-ph.HE]}}.

\bibitem{Amaro-Seoane:2007osp}
P.~Amaro-Seoane, J.~R. Gair, M.~Freitag, M.~Coleman~Miller, I.~Mandel, C.~J.
  Cutler, and S.~Babak, ``{Astrophysics, detection and science applications of
  intermediate- and extreme mass-ratio inspirals},''
  \href{http://dx.doi.org/10.1088/0264-9381/24/17/R01}{{\em Class. Quant.
  Grav.} {\bfseries 24} (2007) R113--R169},
  \href{http://arxiv.org/abs/astro-ph/0703495}{{\ttfamily
  arXiv:astro-ph/0703495}}.

\bibitem{Cardoso:2019rou}
V.~Cardoso and A.~Maselli, ``{Constraints on the astrophysical environment of
  binaries with gravitational-wave observations},''
  \href{http://dx.doi.org/10.1051/0004-6361/202037654}{{\em Astron. Astrophys.}
  {\bfseries 644} (2020) A147},
  \href{http://arxiv.org/abs/1909.05870}{{\ttfamily arXiv:1909.05870
  [astro-ph.HE]}}.

\bibitem{Barausse:2014tra}
E.~Barausse, V.~Cardoso, and P.~Pani, ``{Can environmental effects spoil
  precision gravitational-wave astrophysics?},''
  \href{http://dx.doi.org/10.1103/PhysRevD.89.104059}{{\em Phys. Rev. D}
  {\bfseries 89} no.~10, (2014) 104059},
  \href{http://arxiv.org/abs/1404.7149}{{\ttfamily arXiv:1404.7149 [gr-qc]}}.

\bibitem{Yunes:2011ws}
N.~Yunes, B.~Kocsis, A.~Loeb, and Z.~Haiman, ``{Imprint of Accretion
  Disk-Induced Migration on Gravitational Waves from Extreme Mass Ratio
  Inspirals},'' \href{http://dx.doi.org/10.1103/PhysRevLett.107.171103}{{\em
  Phys. Rev. Lett.} {\bfseries 107} (2011) 171103},
  \href{http://arxiv.org/abs/1103.4609}{{\ttfamily arXiv:1103.4609
  [astro-ph.CO]}}.

\bibitem{Kocsis:2011dr}
B.~Kocsis, N.~Yunes, and A.~Loeb, ``{Observable Signatures of EMRI Black Hole
  Binaries Embedded in Thin Accretion Disks},''
  \href{http://dx.doi.org/10.1103/PhysRevD.86.049907}{{\em Phys. Rev. D}
  {\bfseries 84} (2011) 024032},
  \href{http://arxiv.org/abs/1104.2322}{{\ttfamily arXiv:1104.2322
  [astro-ph.GA]}}.

\bibitem{Destounis:2021mqv}
K.~Destounis, A.~G. Suvorov, and K.~D. Kokkotas, ``{Gravitational-wave glitches
  in chaotic extreme-mass-ratio inspirals},''
  \href{http://dx.doi.org/10.1103/PhysRevLett.126.141102}{{\em Phys. Rev.
  Lett.} {\bfseries 126} no.~14, (2021) 141102},
  \href{http://arxiv.org/abs/2103.05643}{{\ttfamily arXiv:2103.05643 [gr-qc]}}.

\bibitem{Cardoso:2022whc}
V.~Cardoso, K.~Destounis, F.~Duque, R.~Panosso~Macedo, and A.~Maselli,
  ``{Gravitational waves from extreme-mass-ratio systems in astrophysical
  environments},'' \href{http://arxiv.org/abs/2210.01133}{{\ttfamily
  arXiv:2210.01133 [gr-qc]}}.

\bibitem{Cole:2022fir}
P.~S. Cole, G.~Bertone, A.~Coogan, D.~Gaggero, T.~Karydas, B.~J. Kavanagh,
  T.~F.~M. Spieksma, and G.~M. Tomaselli, ``{Disks, spikes, and clouds:
  distinguishing environmental effects on BBH gravitational waveforms},''
  \href{http://arxiv.org/abs/2211.01362}{{\ttfamily arXiv:2211.01362 [gr-qc]}}.

\bibitem{Bamber:2022pbs}
J.~Bamber, J.~C. Aurrekoetxea, K.~Clough, and P.~G. Ferreira, ``{Black hole
  merger simulations in wave dark matter environments},''
  \href{http://arxiv.org/abs/2210.09254}{{\ttfamily arXiv:2210.09254 [gr-qc]}}.

\bibitem{Barack:2018yly}
L.~Barack {\em et~al.}, ``{Black holes, gravitational waves and fundamental
  physics: a roadmap},'' \href{http://dx.doi.org/10.1088/1361-6382/ab0587}{{\em
  Class. Quant. Grav.} {\bfseries 36} no.~14, (2019) 143001},
  \href{http://arxiv.org/abs/1806.05195}{{\ttfamily arXiv:1806.05195 [gr-qc]}}.

\bibitem{Barausse:2020rsu}
E.~Barausse {\em et~al.}, ``{Prospects for Fundamental Physics with LISA},''
  \href{http://dx.doi.org/10.1007/s10714-020-02691-1}{{\em Gen. Rel. Grav.}
  {\bfseries 52} no.~8, (2020) 81},
  \href{http://arxiv.org/abs/2001.09793}{{\ttfamily arXiv:2001.09793 [gr-qc]}}.

\bibitem{Barausse:2016eii}
E.~Barausse, N.~Yunes, and K.~Chamberlain, ``{Theory-Agnostic Constraints on
  Black-Hole Dipole Radiation with Multiband Gravitational-Wave
  Astrophysics},'' \href{http://dx.doi.org/10.1103/PhysRevLett.116.241104}{{\em
  Phys. Rev. Lett.} {\bfseries 116} no.~24, (2016) 241104},
  \href{http://arxiv.org/abs/1603.04075}{{\ttfamily arXiv:1603.04075 [gr-qc]}}.

\bibitem{Blazquez-Salcedo:2016enn}
J.~L. Bl\'azquez-Salcedo, C.~F.~B. Macedo, V.~Cardoso, V.~Ferrari,
  L.~Gualtieri, F.~S. Khoo, J.~Kunz, and P.~Pani, ``{Perturbed black holes in
  Einstein-dilaton-Gauss-Bonnet gravity: Stability, ringdown, and
  gravitational-wave emission},''
  \href{http://dx.doi.org/10.1103/PhysRevD.94.104024}{{\em Phys. Rev. D}
  {\bfseries 94} no.~10, (2016) 104024},
  \href{http://arxiv.org/abs/1609.01286}{{\ttfamily arXiv:1609.01286 [gr-qc]}}.

\bibitem{Glampedakis:2005cf}
K.~Glampedakis and S.~Babak, ``{Mapping spacetimes with LISA: Inspiral of a
  test-body in a `quasi-Kerr' field},''
  \href{http://dx.doi.org/10.1088/0264-9381/23/12/013}{{\em Class. Quant.
  Grav.} {\bfseries 23} (2006) 4167--4188},
  \href{http://arxiv.org/abs/gr-qc/0510057}{{\ttfamily arXiv:gr-qc/0510057}}.

\bibitem{Barack:2006pq}
L.~Barack and C.~Cutler, ``{Using LISA EMRI sources to test off-Kerr deviations
  in the geometry of massive black holes},''
  \href{http://dx.doi.org/10.1103/PhysRevD.75.042003}{{\em Phys. Rev. D}
  {\bfseries 75} (2007) 042003},
  \href{http://arxiv.org/abs/gr-qc/0612029}{{\ttfamily arXiv:gr-qc/0612029}}.

\bibitem{Cardoso:2018zhm}
V.~Cardoso, G.~Castro, and A.~Maselli, ``{Gravitational waves in massive
  gravity theories: waveforms, fluxes and constraints from extreme-mass-ratio
  mergers},'' \href{http://dx.doi.org/10.1103/PhysRevLett.121.251103}{{\em
  Phys. Rev. Lett.} {\bfseries 121} no.~25, (2018) 251103},
  \href{http://arxiv.org/abs/1809.00673}{{\ttfamily arXiv:1809.00673 [gr-qc]}}.

\bibitem{Datta:2019epe}
S.~Datta, R.~Brito, S.~Bose, P.~Pani, and S.~A. Hughes, ``{Tidal heating as a
  discriminator for horizons in extreme mass ratio inspirals},''
  \href{http://dx.doi.org/10.1103/PhysRevD.101.044004}{{\em Phys. Rev. D}
  {\bfseries 101} no.~4, (2020) 044004},
  \href{http://arxiv.org/abs/1910.07841}{{\ttfamily arXiv:1910.07841 [gr-qc]}}.

\bibitem{Pani:2019cyc}
P.~Pani and A.~Maselli, ``{Love in Extrema Ratio},''
  \href{http://dx.doi.org/10.1142/S0218271819440012}{{\em Int. J. Mod. Phys. D}
  {\bfseries 28} no.~14, (2019) 1944001},
  \href{http://arxiv.org/abs/1905.03947}{{\ttfamily arXiv:1905.03947 [gr-qc]}}.

\bibitem{Maggio:2021uge}
E.~Maggio, M.~van~de Meent, and P.~Pani, ``{Extreme mass-ratio inspirals around
  a spinning horizonless compact object},''
  \href{http://dx.doi.org/10.1103/PhysRevD.104.104026}{{\em Phys. Rev. D}
  {\bfseries 104} no.~10, (2021) 104026},
  \href{http://arxiv.org/abs/2106.07195}{{\ttfamily arXiv:2106.07195 [gr-qc]}}.

\bibitem{Destounis:2020kss}
K.~Destounis, A.~G. Suvorov, and K.~D. Kokkotas, ``{Testing spacetime symmetry
  through gravitational waves from extreme-mass-ratio inspirals},''
  \href{http://dx.doi.org/10.1103/PhysRevD.102.064041}{{\em Phys. Rev. D}
  {\bfseries 102} no.~6, (2020) 064041},
  \href{http://arxiv.org/abs/2009.00028}{{\ttfamily arXiv:2009.00028 [gr-qc]}}.

\bibitem{Piovano:2020ooe}
G.~A. Piovano, A.~Maselli, and P.~Pani, ``{Model independent tests of the Kerr
  bound with extreme mass ratio inspirals},''
  \href{http://dx.doi.org/10.1016/j.physletb.2020.135860}{{\em Phys. Lett. B}
  {\bfseries 811} (2020) 135860},
  \href{http://arxiv.org/abs/2003.08448}{{\ttfamily arXiv:2003.08448 [gr-qc]}}.

\bibitem{Annulli:2020ilw}
L.~Annulli, V.~Cardoso, and R.~Vicente, ``{Stirred and shaken: Dynamical
  behavior of boson stars and dark matter cores},''
  \href{http://dx.doi.org/10.1016/j.physletb.2020.135944}{{\em Phys. Lett. B}
  {\bfseries 811} (2020) 135944},
  \href{http://arxiv.org/abs/2007.03700}{{\ttfamily arXiv:2007.03700
  [astro-ph.HE]}}.

\bibitem{Sago:2021iku}
N.~Sago and T.~Tanaka, ``{Oscillations in the extreme mass-ratio inspiral
  gravitational wave phase correction as a probe of a reflective boundary of
  the central black hole},''
  \href{http://dx.doi.org/10.1103/PhysRevD.104.064009}{{\em Phys. Rev. D}
  {\bfseries 104} no.~6, (2021) 064009},
  \href{http://arxiv.org/abs/2106.07123}{{\ttfamily arXiv:2106.07123 [gr-qc]}}.

\bibitem{Piovano:2022ojl}
G.~A. Piovano, A.~Maselli, and P.~Pani, ``{Extreme Love in the SPA:
  constraining the tidal deformability of supermassive objects with extreme
  mass ratio inspirals and semi-analytical, frequency-domain waveforms},''
  \href{http://arxiv.org/abs/2207.07452}{{\ttfamily arXiv:2207.07452 [gr-qc]}}.

\bibitem{Pani:2011xj}
P.~Pani, V.~Cardoso, and L.~Gualtieri, ``{Gravitational waves from extreme
  mass-ratio inspirals in Dynamical Chern-Simons gravity},''
  \href{http://dx.doi.org/10.1103/PhysRevD.83.104048}{{\em Phys. Rev. D}
  {\bfseries 83} (2011) 104048},
  \href{http://arxiv.org/abs/1104.1183}{{\ttfamily arXiv:1104.1183 [gr-qc]}}.

\bibitem{Yunes:2011aa}
N.~Yunes, P.~Pani, and V.~Cardoso, ``{Gravitational Waves from Quasicircular
  Extreme Mass-Ratio Inspirals as Probes of Scalar-Tensor Theories},''
  \href{http://dx.doi.org/10.1103/PhysRevD.85.102003}{{\em Phys. Rev. D}
  {\bfseries 85} (2012) 102003},
  \href{http://arxiv.org/abs/1112.3351}{{\ttfamily arXiv:1112.3351 [gr-qc]}}.

\bibitem{Hannuksela:2018izj}
O.~A. Hannuksela, K.~W.~K. Wong, R.~Brito, E.~Berti, and T.~G.~F. Li,
  ``{Probing the existence of ultralight bosons with a single
  gravitational-wave measurement},''
  \href{http://dx.doi.org/10.1038/s41550-019-0712-4}{{\em Nature Astron.}
  {\bfseries 3} no.~5, (2019) 447--451},
  \href{http://arxiv.org/abs/1804.09659}{{\ttfamily arXiv:1804.09659
  [astro-ph.HE]}}.

\bibitem{Maselli:2020zgv}
A.~Maselli, N.~Franchini, L.~Gualtieri, and T.~P. Sotiriou, ``{Detecting scalar
  fields with Extreme Mass Ratio Inspirals},''
  \href{http://dx.doi.org/10.1103/PhysRevLett.125.141101}{{\em Phys. Rev.
  Lett.} {\bfseries 125} no.~14, (2020) 141101},
  \href{http://arxiv.org/abs/2004.11895}{{\ttfamily arXiv:2004.11895 [gr-qc]}}.

\bibitem{Maselli:2021men}
A.~Maselli, N.~Franchini, L.~Gualtieri, T.~P. Sotiriou, S.~Barsanti, and
  P.~Pani, ``{Detecting fundamental fields with LISA observations of
  gravitational waves from extreme mass-ratio inspirals},''
  \href{http://dx.doi.org/10.1038/s41550-021-01589-5}{{\em Nature Astron.}
  {\bfseries 6} no.~4, (2022) 464--470},
  \href{http://arxiv.org/abs/2106.11325}{{\ttfamily arXiv:2106.11325 [gr-qc]}}.

\bibitem{Collodel:2021jwi}
L.~G. Collodel, D.~D. Doneva, and S.~S. Yazadjiev, ``{Equatorial
  extreme-mass-ratio inspirals in Kerr black holes with scalar hair
  spacetimes},'' \href{http://dx.doi.org/10.1103/PhysRevD.105.044036}{{\em
  Phys. Rev. D} {\bfseries 105} no.~4, (2022) 044036},
  \href{http://arxiv.org/abs/2108.11658}{{\ttfamily arXiv:2108.11658 [gr-qc]}}.

\bibitem{Barsanti:2022ana}
S.~Barsanti, N.~Franchini, L.~Gualtieri, A.~Maselli, and T.~P. Sotiriou,
  ``{Extreme mass-ratio inspirals as probes of scalar fields: Eccentric
  equatorial orbits around Kerr black holes},''
  \href{http://dx.doi.org/10.1103/PhysRevD.106.044029}{{\em Phys. Rev. D}
  {\bfseries 106} no.~4, (2022) 044029},
  \href{http://arxiv.org/abs/2203.05003}{{\ttfamily arXiv:2203.05003 [gr-qc]}}.

\bibitem{Berti:2015itd}
E.~Berti {\em et~al.}, ``{Testing General Relativity with Present and Future
  Astrophysical Observations},''
  \href{http://dx.doi.org/10.1088/0264-9381/32/24/243001}{{\em Class. Quant.
  Grav.} {\bfseries 32} (2015) 243001},
  \href{http://arxiv.org/abs/1501.07274}{{\ttfamily arXiv:1501.07274 [gr-qc]}}.

\bibitem{Brito:2015oca}
R.~Brito, V.~Cardoso, and P.~Pani, ``{Superradiance}: {New Frontiers in Black
  Hole Physics},'' \href{http://dx.doi.org/10.1007/978-3-319-19000-6}{{\em
  Lect. Notes Phys.} {\bfseries 906} (2015) pp.1--237},
  \href{http://arxiv.org/abs/1501.06570}{{\ttfamily arXiv:1501.06570 [gr-qc]}}.

\bibitem{Brito:2017zvb}
R.~Brito, S.~Ghosh, E.~Barausse, E.~Berti, V.~Cardoso, I.~Dvorkin, A.~Klein,
  and P.~Pani, ``{Gravitational wave searches for ultralight bosons with LIGO
  and LISA},'' \href{http://dx.doi.org/10.1103/PhysRevD.96.064050}{{\em Phys.
  Rev. D} {\bfseries 96} no.~6, (2017) 064050},
  \href{http://arxiv.org/abs/1706.06311}{{\ttfamily arXiv:1706.06311 [gr-qc]}}.

\bibitem{Damour:1993hw}
T.~Damour and G.~Esposito-Farese, ``{Nonperturbative strong field effects in
  tensor - scalar theories of gravitation},''
  \href{http://dx.doi.org/10.1103/PhysRevLett.70.2220}{{\em Phys. Rev. Lett.}
  {\bfseries 70} (1993) 2220--2223}.

\bibitem{Silva:2017uqg}
H.~O. Silva, J.~Sakstein, L.~Gualtieri, T.~P. Sotiriou, and E.~Berti,
  ``{Spontaneous scalarization of black holes and compact stars from a
  Gauss-Bonnet coupling},''
  \href{http://dx.doi.org/10.1103/PhysRevLett.120.131104}{{\em Phys. Rev.
  Lett.} {\bfseries 120} no.~13, (2018) 131104},
  \href{http://arxiv.org/abs/1711.02080}{{\ttfamily arXiv:1711.02080 [gr-qc]}}.

\bibitem{Doneva:2017bvd}
D.~D. Doneva and S.~S. Yazadjiev, ``{New Gauss-Bonnet Black Holes with
  Curvature-Induced Scalarization in Extended Scalar-Tensor Theories},''
  \href{http://dx.doi.org/10.1103/PhysRevLett.120.131103}{{\em Phys. Rev.
  Lett.} {\bfseries 120} no.~13, (2018) 131103},
  \href{http://arxiv.org/abs/1711.01187}{{\ttfamily arXiv:1711.01187 [gr-qc]}}.

\bibitem{Dima:2020yac}
A.~Dima, E.~Barausse, N.~Franchini, and T.~P. Sotiriou, ``{Spin-induced black
  hole spontaneous scalarization},''
  \href{http://dx.doi.org/10.1103/PhysRevLett.125.231101}{{\em Phys. Rev.
  Lett.} {\bfseries 125} no.~23, (2020) 231101},
  \href{http://arxiv.org/abs/2006.03095}{{\ttfamily arXiv:2006.03095 [gr-qc]}}.

\bibitem{Doneva:2022ewd}
D.~D. Doneva, F.~M. Ramazano\u{g}lu, H.~O. Silva, T.~P. Sotiriou, and S.~S.
  Yazadjiev, ``{Scalarization},''
  \href{http://arxiv.org/abs/2211.01766}{{\ttfamily arXiv:2211.01766 [gr-qc]}}.

\bibitem{Antoniadis:2013pzd}
J.~Antoniadis {\em et~al.}, ``{A Massive Pulsar in a Compact Relativistic
  Binary},'' \href{http://dx.doi.org/10.1126/science.1233232}{{\em Science}
  {\bfseries 340} (2013) 6131},
  \href{http://arxiv.org/abs/1304.6875}{{\ttfamily arXiv:1304.6875
  [astro-ph.HE]}}.

\bibitem{Yamada:2019zrb}
K.~Yamada, T.~Narikawa, and T.~Tanaka, ``{Testing massive-field modifications
  of gravity via gravitational waves},''
  \href{http://dx.doi.org/10.1093/ptep/ptz103}{{\em PTEP} {\bfseries 2019}
  no.~10, (2019) 103E01}, \href{http://arxiv.org/abs/1905.11859}{{\ttfamily
  arXiv:1905.11859 [gr-qc]}}.

\bibitem{Ramazanoglu:2016kul}
F.~M. Ramazano\u{g}lu and F.~Pretorius, ``{Spontaneous Scalarization with
  Massive Fields},'' \href{http://dx.doi.org/10.1103/PhysRevD.93.064005}{{\em
  Phys. Rev. D} {\bfseries 93} no.~6, (2016) 064005},
  \href{http://arxiv.org/abs/1601.07475}{{\ttfamily arXiv:1601.07475 [gr-qc]}}.

\bibitem{1975ApJ...196L..59E}
D.~M. {Eardley}, ``{Observable effects of a scalar gravitational field in a
  binary pulsar.},'' \href{http://dx.doi.org/10.1086/181744}{{\em The
  Astrophysical Journal} {\bfseries 196} (Mar, 1975) L59--L62}.

\bibitem{Nair:2019iur}
R.~Nair, S.~Perkins, H.~O. Silva, and N.~Yunes, ``{Fundamental Physics
  Implications for Higher-Curvature Theories from Binary Black Hole Signals in
  the LIGO-Virgo Catalog GWTC-1},''
  \href{http://dx.doi.org/10.1103/PhysRevLett.123.191101}{{\em Phys. Rev.
  Lett.} {\bfseries 123} no.~19, (2019) 191101},
  \href{http://arxiv.org/abs/1905.00870}{{\ttfamily arXiv:1905.00870 [gr-qc]}}.

\bibitem{Bekenstein:1995un}
J.~D. Bekenstein, ``{Novel
  \textquoteleft{}\textquoteleft{}no-scalar-hair\textquoteright{}\textquoteright{}
  theorem for black holes},''
  \href{http://dx.doi.org/10.1103/PhysRevD.51.R6608}{{\em Phys. Rev. D}
  {\bfseries 51} no.~12, (1995) R6608}.

\bibitem{Sotiriou:2011dz}
T.~P. Sotiriou and V.~Faraoni, ``{Black holes in scalar-tensor gravity},''
  \href{http://dx.doi.org/10.1103/PhysRevLett.108.081103}{{\em Phys. Rev.
  Lett.} {\bfseries 108} (2012) 081103},
  \href{http://arxiv.org/abs/1109.6324}{{\ttfamily arXiv:1109.6324 [gr-qc]}}.

\bibitem{Teukolsky:1973ha}
S.~A. Teukolsky, ``{Perturbations of a rotating black hole. 1. Fundamental
  equations for gravitational electromagnetic and neutrino field
  perturbations},'' \href{http://dx.doi.org/10.1086/152444}{{\em Astrophys. J.}
  {\bfseries 185} (1973) 635--647}.

\bibitem{Alsing:2011er}
J.~Alsing, E.~Berti, C.~M. Will, and H.~Zaglauer, ``{Gravitational radiation
  from compact binary systems in the massive Brans-Dicke theory of gravity},''
  \href{http://dx.doi.org/10.1103/PhysRevD.85.064041}{{\em Phys. Rev. D}
  {\bfseries 85} (2012) 064041},
  \href{http://arxiv.org/abs/1112.4903}{{\ttfamily arXiv:1112.4903 [gr-qc]}}.

\bibitem{Berti:2012bp}
E.~Berti, L.~Gualtieri, M.~Horbatsch, and J.~Alsing, ``{Light scalar field
  constraints from gravitational-wave observations of compact binaries},''
  \href{http://dx.doi.org/10.1103/PhysRevD.85.122005}{{\em Phys. Rev. D}
  {\bfseries 85} (2012) 122005},
  \href{http://arxiv.org/abs/1204.4340}{{\ttfamily arXiv:1204.4340 [gr-qc]}}.

\bibitem{Cardoso:2011xi}
V.~Cardoso, S.~Chakrabarti, P.~Pani, E.~Berti, and L.~Gualtieri, ``{Floating
  and sinking: The Imprint of massive scalars around rotating black holes},''
  \href{http://dx.doi.org/10.1103/PhysRevLett.107.241101}{{\em Phys. Rev.
  Lett.} {\bfseries 107} (2011) 241101},
  \href{http://arxiv.org/abs/1109.6021}{{\ttfamily arXiv:1109.6021 [gr-qc]}}.

\bibitem{BHPToolkit}
``{Black Hole Perturbation Toolkit}.''
  (\href{http://bhptoolkit.org/}{bhptoolkit.org}).

\bibitem{SGREP_REPO}
``{sgrep repo}.''
  (\href{https://github.com/masellia/SGREP/}{github.com/masellia/SGREP/)},.

\bibitem{Robson:2018ifk}
T.~Robson, N.~J. Cornish, and C.~Liu, ``{The construction and use of LISA
  sensitivity curves},'' \href{http://dx.doi.org/10.1088/1361-6382/ab1101}{{\em
  Class. Quant. Grav.} {\bfseries 36} no.~10, (2019) 105011},
  \href{http://arxiv.org/abs/1803.01944}{{\ttfamily arXiv:1803.01944
  [astro-ph.HE]}}.

\bibitem{Chatziioannou:2017tdw}
K.~Chatziioannou, A.~Klein, N.~Yunes, and N.~Cornish, ``{Constructing
  Gravitational Waves from Generic Spin-Precessing Compact Binary Inspirals},''
  \href{http://dx.doi.org/10.1103/PhysRevD.95.104004}{{\em Phys. Rev. D}
  {\bfseries 95} no.~10, (2017) 104004},
  \href{http://arxiv.org/abs/1703.03967}{{\ttfamily arXiv:1703.03967 [gr-qc]}}.

\bibitem{Babak:2017tow}
S.~Babak, J.~Gair, A.~Sesana, E.~Barausse, C.~F. Sopuerta, C.~P.~L. Berry,
  E.~Berti, P.~Amaro-Seoane, A.~Petiteau, and A.~Klein, ``{Science with the
  space-based interferometer LISA. V: Extreme mass-ratio inspirals},''
  \href{http://dx.doi.org/10.1103/PhysRevD.95.103012}{{\em Phys. Rev. D}
  {\bfseries 95} no.~10, (2017) 103012},
  \href{http://arxiv.org/abs/1703.09722}{{\ttfamily arXiv:1703.09722 [gr-qc]}}.

\bibitem{papeinclined}
M.~Della~Rocca {\em et~al.}, ``in preparation,''.

\bibitem{Zhang:2022rfr}
C.~Zhang, Y.~Gong, D.~Liang, and B.~Wang, ``{Gravitational waves from eccentric
  extreme mass-ratio inspirals as probes of scalar fields},''
  \href{http://arxiv.org/abs/2210.11121}{{\ttfamily arXiv:2210.11121 [gr-qc]}}.

\bibitem{sperinew}
L.~Speri and et~al., ``{Fast Emri Waveforms to test General Relativity},''
  \href{http://arxiv.org/abs/in preparation}{{\ttfamily in preparation}}.

\bibitem{Bonga:2019ycj}
B.~Bonga, H.~Yang, and S.~A. Hughes, ``{Tidal resonance in extreme mass-ratio
  inspirals},'' \href{http://dx.doi.org/10.1103/PhysRevLett.123.101103}{{\em
  Phys. Rev. Lett.} {\bfseries 123} no.~10, (2019) 101103},
  \href{http://arxiv.org/abs/1905.00030}{{\ttfamily arXiv:1905.00030 [gr-qc]}}.

\end{thebibliography}\endgroup

\end{document}